\documentstyle[psfig]{mn}

\newif\ifAMStwofonts

\def\simlt{\lower.5ex\hbox{$\; \buildrel < \over \sim \;$}}
\def\simgt{\lower.5ex\hbox{$\; \buildrel > \over \sim \;$}}

\ifoldfss
  \ifCUPmtlplainloaded \else
    \NewTextAlphabet{textbfit} {cmbxti10} {}
    \NewTextAlphabet{textbfss} {cmssbx10} {}
    \NewMathAlphabet{mathbfit} {cmbxti10} {} 
    \NewMathAlphabet{mathbfss} {cmssbx10} {} 
  \fi
  \ifAMStwofonts
    \ifCUPmtlplainloaded \else
      \NewSymbolFont{upmath} {eurm10}
      \NewSymbolFont{AMSa} {msam10}
      \NewMathSymbol{\upi}     {0}{upmath}{19}
      \NewMathSymbol{\umu}     {0}{upmath}{16}
      \NewMathSymbol{\upartial}{0}{upmath}{40}
      \NewMathSymbol{\leqslant}{3}{AMSa}{36}
      \NewMathSymbol{\geqslant}{3}{AMSa}{3E}

       \let\le=\leqslant
       
    \fi
  \fi
\fi 

\ifnfssone
  \newmathalphabet{\mathit}
  \addtoversion{normal}{\mathit}{cmr}{m}{it}
  \addtoversion{bold}{\mathit}{cmr}{bx}{it}
  \newmathalphabet{\mathbfit} 
  \addtoversion{normal}{\mathbfit}{cmr}{bx}{it}
  \addtoversion{bold}{\mathbfit}{cmr}{bx}{it}
  \newmathalphabet{\mathbfss} 
  \addtoversion{normal}{\mathbfss}{cmss}{bx}{n}
  \addtoversion{bold}{\mathbfss}{cmss}{bx}{n}
  \ifAMStwofonts
    \ifCUPmtlplainloaded \else
      \UseAMStwoboldmath
      \makeatletter
      \new@mathgroup\upmath@group
      \define@mathgroup\mv@normal\upmath@group{eur}{m}{n}
      \define@mathgroup\mv@bold\upmath@group{eur}{b}{n}
      \edef\UPM{\hexnumber\upmath@group}
      \new@mathgroup\amsa@group
      \define@mathgroup\mv@normal\amsa@group{msa}{m}{n}
      \define@mathgroup\mv@bold\amsa@group{msa}{m}{n}
      \edef\AMSa{\hexnumber\amsa@group}
      \makeatother
      \mathchardef\upi="0\UPM19
      \mathchardef\umu="0\UPM16
      \mathchardef\upartial="0\UPM40
      \mathchardef\leqslant="3\AMSa36
      \mathchardef\geqslant="3\AMSa3E

       \let\le=\leqslant

    \fi
  \fi
\fi 

\ifnfsstwo
  \DeclareMathAlphabet{\mathbfit}{OT1}{cmr}{bx}{it}
  \SetMathAlphabet\mathbfit{bold}{OT1}{cmr}{bx}{it}
  \DeclareMathAlphabet{\mathbfss}{OT1}{cmss}{bx}{n}
  \SetMathAlphabet\mathbfss{bold}{OT1}{cmss}{bx}{n}
  \ifAMStwofonts
    \ifCUPmtlplainloaded \else
      \DeclareSymbolFont{UPM}{U}{eur}{m}{n}
      \SetSymbolFont{UPM}{bold}{U}{eur}{b}{n}
      \DeclareSymbolFont{AMSa}{U}{msa}{m}{n}
      \DeclareMathSymbol{\upi}{0}{UPM}{"19}
      \DeclareMathSymbol{\umu}{0}{UPM}{"16}
      \DeclareMathSymbol{\upartial}{0}{UPM}{"40}
      \DeclareMathSymbol{\leqslant}{3}{AMSa}{"36}
      \DeclareMathSymbol{\geqslant}{3}{AMSa}{"3E}

       \let\le=\leqslant

    \fi
  \fi
\fi 

\ifCUPmtlplainloaded \else
  \ifAMStwofonts \else 
    \def\upi{\pi}
    \def\umu{\mu}
    \def\upartial{\partial}
  \fi
\fi

\title[The Evolution of Substructure]
    {The Evolution of Substructure in Galaxy, \\ Group and Cluster Haloes I:
Basic Dynamics}
\author[J.E.\ Taylor and A.\ Babul]
{James E. Taylor$^{1}$\thanks{email: {\tt jet@astro.ox.ac.uk}} and Arif Babul$^{2}$\thanks{CITA Senior Fellow} \\
$^{1}$Denys Wilkinson Building, 1 Keble Road, Oxford OX1 3RH, United Kingdom \\
$^{2}$Elliott Building, 3800 Finnerty Road, Victoria, BC, V8P 1A1, Canada \\}
\date{\today}
\pubyear{2002}

\begin{document}

\maketitle

\begin{abstract}
The hierarchical mergers that form the haloes of dark matter surrounding
galaxies, groups and clusters are not entirely efficient, leaving 
substantial amounts of dense substructure, in the form of stripped halo 
cores or `subhaloes', orbiting within these systems.
Using a semi-analytic model of satellite dynamics, we study the
evolution of haloes as they merge hierarchically, to determine how 
much substructure survives merging and how the properties 
of individual subhaloes change over time. 
We find that subhaloes evolve, due to mass loss, orbital 
decay, and tidal disruption, on a characteristic time-scale equal to 
the period of radial oscillations at the virial radius of the system.
Subhaloes with realistic
densities and density profiles lose 25--45 per cent of their mass per 
pericentric passage, depending on their concentration and 
on the circularity of their orbit.
As the halo grows, the subhalo orbits also grow in size and become less 
bound. Based on these general patterns, we suggest a method for including 
realistic amounts of substructure in semi-analytic models based on merger 
trees. We show that the parameters in the resulting model can be 
fixed by requiring self-consistency between different levels of the 
merger hierarchy. In a companion paper, we will compare 
the results of our model with numerical simulations of halo formation.
\end{abstract}


\section{Introduction}\label{sec:intro}

In the standard picture of structure formation in a universe 
dominated by cold dark matter (CDM), small fluctuations present 
in the density field at early times grow through gravitational 
instability. Eventually, when their amplitude is large enough, 
they cease expanding with the Hubble flow, collapse and virialise,
forming dense, relaxed systems, or `haloes'. Dark matter
haloes are important as sites of galaxy formation (White \& Rees 1978) 
and of the subsequent formation of groups and clusters 
through hierarchical merging (Blumenthal et al.\ 1984). As 
the densest concentrations of dark matter, they are also the best 
places to search for evidence of decays of dark matter particles 
(e.g.\ Gondolo \& Silk 1999; Blasi \& Seth 2001; see Bergstr\"{o}m 2000 
for a recent review), (self-)interactions 
(e.g.\ Peebles \& Vilenkin 1999; Spergel \& Steinhardt 2000; 
see Natarajan et al.\ 2002 for recent observational limits in clusters), 
or other dark matter physics such as bosonic or scalar properties, 
(e.g.\ Goodman 2000; Hu, Barkana, \& Gruzinov 2000) connections to 
quintessence (e.g.\ Wetterich 2002; Padmanabhan \& Choudhury 2002), 
interactions with photons (e.g.\ B{\oe}hm et al.\ 2002), or 
interactions with baryons 
(e.g.\ Cyburt et al.\ 2002). Finally, since our own galaxy
should be embedded in a dark matter halo, a detailed understanding 
of halo substructure is essential to interpreting the limits placed 
by experiments to detect dark matter directly in the solar neighbourhood
(e.g.\ Helmi, White, \& Springel 2002; see Pretzl 2002 for a recent review).

Numerical simulations of structure formation predict a generic 
density profile for dark matter haloes (Navarro, Frenk \& White 1996, 1997; 
Moore et al.\ 1998) that is in rough agreement with those inferred from 
X-ray and lensing observations 
of galaxy clusters, at least in their outer regions (e.g.\ David et al.\ 2001;
Arabadjis, Bautz, \& Garmire 2002; Sand, Treu \& Ellis 2002; 
Lewis, Buote \& Stocke 2002) as well as from recent weak-lensing studies 
(e.g.\ Hoekstra et al.\ 2002). For less massive haloes the agreement is less 
certain (e.g.\ Blais-Ouellette, Amram, \& Carignan 2001; 
Borriello \& Salucci 2001; de Blok \& Bosma 2002; Marchesini et al.\ 2002 
and earlier references therein), 
but on these scales the physics of galaxy formation may have played a greater 
role in rearranging material within haloes. 
Recently, the analysis of multiply-lensed quasars (Chiba 2002;
Metcalf and Zhao 2002; 
Brada{\v c} et al.\ 2002; Dalal \& Kochanek 2002; Keeton 2002) and of
bending in the images of radio jets (Metcalf 2002) has also provided 
evidence that dark 
matter haloes contain a substantial amount of dense substructure. 
In the highest-resolution simulations, this substructure is seen to
result from the relative inefficiency of the hierarchical
merging process: when haloes merge together, their dense cores can
often survive for many orbits in the resulting system, 
as distinct, self-bound substructure.
The overall number and mass distribution of these cores, or `subhaloes',
is predicted to have a universal form, roughly independent of halo mass,
over many orders of magnitude, both in SCDM 
(Klypin et al.\ 1999; Okamoto \& Habe 1999; 
Moore et al.\ 1999; Ghigna et al.\ 2000) and in LCDM (Springel et al.\ 2001; 
Font et al.\ 2001; Governato et al.\ 2001; Stoehr et al 2002) cosmologies. 

Although substructure accounts for only 10 per cent of the mass of an 
average system, because of its high density it has important implications both 
for galaxy formation and for models of dark matter physics. Substructure can 
disrupt fragile structures in galaxy haloes, such as tidal streams 
(Johnston, Spergel, \& Haydn 2002; Ibata et al.\ 2002; Mayer et al.\ 2002) 
or galactic disks (T\'{o}th \& Ostriker 1992; see Taylor \& Babul 2001 for 
more recent references). It may substantially enhance the 
rate of interactions between dark matter particles, increasing the 
annihilation signal from dark matter if it consists of WIMPS 
(Calc{\' a}neo-Rold{\' a}n \& Moore 2000; Ullio et al.\ 2002; 
Taylor \& Silk 2002). It may also account for the large mass-to-light ratios
measured in Local Group satellites (Mateo 1998), as well as explaining 
the multiple-lensing results discussed previously. 

To develop tests of the nature of dark matter at high densities and on 
small spatial scales where its properties may be most obvious, and to 
predict halo structure and substructure for galaxy and cluster formation 
models, halo substructure must be modelled accurately down to extremely small 
mass scales. Weak lensing, for instance, may be sensitive to substructure 
on scales of
$10^{3} {\rm M}_{\odot}$ (Metcalf \& Madau 2001), that is $10^{-9}$ times 
the mass of a galaxy halo. Unfortunately, the mass resolution limit in current
numerical simulations of halo formation is much larger than this. Scaled to 
the halo of the Milky Way, the highest-resolution simulations of haloes 
presently available have particle masses of a few times $10^{5} {\rm M}_{\odot}$ 
(Springel et al.\ 2001), and thus they can only resolve halo properties 
down to $\simeq 10^{7} {\rm M}_{\odot}$, four orders of magnitude larger than 
the scale required. Clearly analytic or semi-analytic techniques are required 
to extend the predictions of hierarchical models any further in the near 
future. 

In a previous paper (Taylor \& Babul 2001, TB01 hereafter), we developed
an analytic model for the dynamical evolution of satellites orbiting
in the potential of a larger system. This model includes a simplified
description of dynamical friction and of mass loss due to tidal truncation 
and tidal heating, using a set of evolution equations based on the global 
properties of a satellite to modify its mass, structure and orbit over
a short time-step, rather like a restricted $N$-body simulation. As a result, 
it is well suited to inclusion in semi-analytic models, which must 
follow the evolution of large numbers of systems over many time-steps 
without excessive computational cost. In this paper, we extend the work
in TB01 to a full model of halo formation
based on semi-analytic merger trees.

A key problem we will address in this paper is the treatment of higher-order 
substructure, that is the substructure already present in subsidiary 
haloes when they merge with the main progenitor of a system. The cores
of subhaloes are sufficiently robust that they may survive through
many levels of the merger hierarchy, contributing to the subhalo
population of the final system. In previous semi-analytic models of 
halo substructure, this contribution has either been ignored, by 
treating all haloes merging with the main
progenitor as single objects with no substructure 
(e.g.\ Bullock, Kravtsov, \& Weinberg 2000, 2001a), or the merger trees have
been `pruned' of higher-order substructure by assuming that it
merges on the dynamical friction time-scale (e.g.\ Somerville 2002). 
Only the model of Benson et al.\ (2002a, 2002b) follows higher-order
substructure in detail, by integrating orbits and calculating mass
loss due to tidal stripping over many time-steps. 
As we will show in this paper, none of these
approaches are entirely satisfactory: higher-order substructure
should contribute substantially to the subhalo mass function, so it cannot
simply be ignored; dynamical friction is largely ineffective for
satellites with less than $10^{-2}$--$10^{-3}$ of the mass of the main
system (e.g.\ Taffoni et al.\ 2002), so it will not prune merger trees 
efficiently below 
this level; and following the evolution of substructure in detail at every 
level of a merger tree greatly reduces the speed of semi-analytic 
calculations.

In this paper, we propose a simple method for reducing the complexity of
a merger tree, based on the patterns that appear when we follow the
dynamical evolution of systems in the main branch of the merger tree 
in detail. Haloes falling into a larger system should lose their bound 
substructure as they lose mass. We use the spatial distribution of subhaloes 
within a system to calculate the correspondence between the relative rates 
for these two processes. We assume that the most recently acquired substructure
is the first to be stripped off, since it is typically on more extended and
less bound orbits. This gives us a precise criterion for pruning the
merger tree, passing substructure on to the next level of the tree if
it has spent less than a certain number of orbits in the halo of its parent.
The parameters of our method are then fixed by requiring self-consistency, 
that is the average mass-loss rates assumed when pruning
the merger trees are the same as those measured for satellites in the
main halo. As a result, the method has no major free parameters 
beyond those introduced in TB01 to follow the evolution of individual 
satellites.

The outline of this paper is as follows. In section \ref{sec:basicmodel},
we review merger-tree models and explain how we establish the
basic conditions for studying satellite evolution in a hierarchical setting. 
In section 
\ref{sec:static}, we examine the behaviour of realistic
satellites in a static halo with fixed properties, using the model of TB01. 
In section 
\ref{sec:dynamic}, we then show how the evolution of satellites
changes when we take into account the changes in halo mass, size and density 
profile predicted in section \ref{sec:basicmodel}.
In section \ref{sec:pruning}, we outline a simple method for pruning
higher-order substructure in merger trees based on these results. We summarise
our results in section \ref{sec:summary}.
In a subsequent paper (Taylor \& Babul in preparation, paper II hereafter), 
we will compare 
the predictions of this model with the results of high-resolution 
simulations, study dynamical groups within halo substructure, 
and discuss the overall evolution of substructure with time. 

Throughout this paper we consider results for a `standard'
CDM (SCDM) cosmology with $\Omega = 1$, $\Lambda = 0$, 
$H_0 = h$ 100 km\,s${^{-1}}$ where $h = 0.5$, $\sigma_8 = 0.7$ 
and $\Gamma = 0.5$, unless otherwise noted, since our primary
goal is to compare our results to simulations in this cosmology. 
In general, we expect very similar results for LCDM, which
produces a similar subhalo mass function (albeit with indications of
a lower normalisation -- cf.\ Springel et al.\ 2001; 
Font et al.\ 2001; Governato et al.\ 2001; Stoehr et al 2002), 
or other CDM cosmologies. 
Furthermore, since 
our method is self-calibrating, it can be used for this or other 
cosmologies without reference to simulations.

\section{Modelling halo formation}\label{sec:basicmodel}

The growth of dark matter haloes through accretion and mergers can be 
predicted approximately using simple statistical arguments, the so-called
`extended Press-Schechter' (EPS) formalism 
(Press \& Schechter 1974; Bower 1991; Bond et al.\ 1991; Lacey \& Cole 1993
-- LC93 hereafter). 
The resulting analytic expressions form the basis 
of many of the semi-analytic models of galaxy formation that have been 
developed extensively over the past decade 
(Kauffmann, White \& Guiderdoni 1993; Cole et al.\ 1994, 2000; 
Somerville \& Primack 1999; see Somerville \& Primack 1999 
and Hatton et al.\ 2003 for recent 
reviews). 
In this section we will review how these methods can be used to 
model the growth of dark matter haloes through hierarchical merging. 
We will also summarise the parameters we have used to generate the 
merger trees 
considered in this paper and in paper II, and test their properties.
We will then explain how the evolution of halo structural properties,
mergers with other haloes, and the evolution of halo substructure 
can be included within this framework.

\subsection{Determining halo growth rates}\label{subsec:2.1}

One can estimate the rate at which virialised haloes form and grow 
by considering the evolution of a spherically symmetric overdensity 
with some small initial amplitude (e.g.\ Peebles 1980). Multiplying 
the initial, linear growth rate by the time at which the fully 
non-linear solution collapses to zero radius, one finds that the
collapse occurs at the epoch when the region reaches a linearly-extrapolated 
overdensity $\delta \equiv (\rho - \rho_{\rm c})/\rho_{\rm c} = 1.686$ 
for the SCDM cosmology considered here. We will refer to this value as
the `critical' overdensity, $\delta_{\rm c}$, hereafter.

This leads to a particularly simple method for identifying 
regions that will virialise at some later time -- the Press-Schechter 
approach (Press \& Schechter 1974). If one considers the density
field at early times when fluctuations are small, then the spherical
collapse model predicts that a region will have collapsed and virialised
by the epoch $z$ if $\delta$, its mean overdensity extrapolated linearly to
the present, satisfies
\begin{equation}
\delta > \delta_{\rm c}/D(z)\,,
\label{eq:collapse}
\end{equation}
where $D(z)$ is the linear growth factor relative to the present day
and $\delta_{\rm c}$ is the critical overdensity. 
Thus in principle, one can construct a history of the
virialised mass around any given point (or `mass accretion history') 
by filtering the density field around that point on different scales, 
and finding the largest scale 
that satisfies equation (\ref{eq:collapse}) at any given epoch. In practice,
for an initial density field with Gaussian fluctuations, the
statistics of mass accretion histories are sufficiently simple
that they can be described by analytic expressions 
(Bower 1991; Bond et al.\ 1991; LC93). 
Applying Monte-Carlo methods, one can use these expressions to generate 
statistically representative realisations of $M(z)$ for individual haloes, 
for a given power spectrum of initial fluctuations (which determines likely
values of $\delta$ on different scales) and a given cosmology
(which determines $\delta_{\rm c}$ the growth factor $D(z)$).

\subsubsection{Generating merger trees}

If one assumes that sudden jumps\footnote{That is to say changes in mass over
a single time-step that exceed the mass resolution of the accretion history;
changes below this limit are treated as smooth accretion.}  
in the virialised mass of any given halo
correspond to mergers with other individual haloes, then the mass accretion 
history will provide a list of all the mergers which contributed to the
main progenitor of the final system. By repeatedly generating mass accretion 
histories for each halo merging with the main progenitor, and then for each 
halo merging with each of these systems, one can start to draw a merger `tree' 
with a main trunk corresponding to the principal halo, branches corresponding 
to `first-order' mergers with the principal halo, `second-order' branches
off the first-order ones, and so on. 

This approach was developed by several different groups  
(Kauffmann \& White 1993; Lacey \& Cole 1994; Sheth \& Lemson 1999; 
Somerville \& Kolatt 1999, SK99 hereafter), and forms the basis of 
a whole set of semi-analytic models of galaxy formation and halo clustering. 
Although conceptually simple, it can be difficult to implement in 
a way that preserves the statistically properties of halo
mergers. Specifically, the analytic expressions for merger probabilities
used to generate merger histories deal only with binary mergers,
and it is difficult to derive from these expressions an algorithm
that conserves mass within a given tree. We will not discuss this 
problem any further here, but refer the
reader to an excellent discussion of these issues in SK99. To minimise
the need for higher-order merger probabilities, SK99 generate trees
with a step in $\omega \equiv \delta_{\rm c}/D(z)$ 
(or equivalently in $z$) 
which is so small that almost all mergers are binary. Using this model 
they demonstrate very good agreement with (analytic) EPS statistics, as
well as with simulations (Somerville et al.\ 2000).
We have adopted their algorithm to generate the merger
trees considered here. Specifically, when determining
the branching of a given section of the merger tree, we use a 
step in the density threshold:
\begin{equation}
\Delta\omega = (a\,\log_{10} (M_{\rm h}/M_{\rm l}) +b)\Delta\omega_0\,,
\end{equation}
where $M_{\rm h}$ is the mass of the branching halo, $M_{\rm l}$ is the 
resolution limit of the tree, $\Delta \omega = \Delta (\delta_{\rm c}/D(z))$ 
is the present-day critical density threshold for collapse, extrapolated 
linearly back to $z$, 
and the basic step size $\Delta\omega_0$ is chosen so that:
\begin{equation}
\Delta\omega_0 \le \sqrt{\left.{{{\rm d}S}\over{{\rm d}M}}\right|_{M_{\rm h}}\,M_{\rm l}}\,,
\end{equation}
were $S = S(M) = \sigma^2(M)$ is the square of the variance which 
characterises the spectrum of density fluctuations (see SK99, section 6).
We discuss the choice of the constants $a$ and $b$ below.

\subsubsection{Parameter choices and tests}

In this paper and in paper II, we will consider a set of merger
trees designed to trace the merger history of a system like the
present-day Milky Way. As mentioned previously, we assume a SCDM
cosmology in order to compare our results with a specific set of 
simulations. (We will describe the simulations in detail in paper II.) 
The final mass of the present-day system in each tree 
is $M_{\rm vir,0} = 1.6\times10^{12}\,{\rm M}_{\odot}$, 
in the range of current 
estimates for the total mass of the halo of the Milky Way 
of $\simeq 1$--$2\times10^{12}{\rm M}_{\odot}$ 
(e.g.\ Klypin, Zhao, \& Somerville 2002).
We follow the merger history down to a limiting mass resolution of 
$M_{\rm l} = 5\times 10^{7}{\rm M}_{\odot}$, or $3.2\times 10^{-4}$, 
which is comparable to the mass scale of the smallest resolved structures
in the simulations.
(To avoid spurious resolution effects at this boundary, we actually 
follow the decomposition of haloes above this mass limit 
into progenitors with masses as small as $2.5\times 10^{6}{\rm M}_{\odot}$, 
but we do not follow the merger histories of haloes with masses
less than $M_{\rm l}$
any further, so our results become incomplete below this mass.) 
For this choice of parameters, most of the mass in the main system
is added through mergers over the resolution limit, though the method
of SK99 also accounts for accretion below this limit consistently.
The merger histories are traced back to $z = 30$,
although most branches of the tree drop below the mass resolution limit
well before this redshift. 

\begin{figure}
  \centerline{\psfig{figure=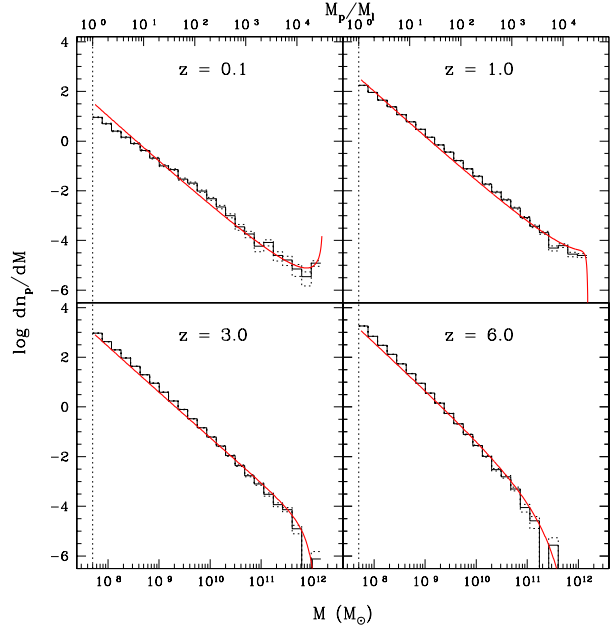,width=1.0\linewidth,clip=,angle=0}}
  \caption[]{The average number of progenitors in a merger tree as
a function of mass (in $M_{\odot}$ along the bottom axis, or in units
of the resolution limit $M_{\rm l}$ along the top axis), at four different 
redshifts (histogram). The dotted histograms show the error in the mean,
and the smooth curves show the Press-Schechter predictions.}
\label{nfig:1}
\end{figure}

The choice of the time-step scaling parameters $a$ and $b$ 
in the SK99 algorithm requires some experimentation. 
Large time-steps will reduce the statistical accuracy of the approach
by producing large numbers of multiple mergers, while very short
time-steps increase the time required to generate each tree,
and may also affect the merger statistics through roundoff
errors and other more complicated effects. We have chosen the values
$a = 0.2$ and $b = 0.1$, which we find to give good results with a 
reasonable computation time per tree for the values 
$M_0 = 1.6\times10^{12}{\rm M}_{\odot}$ and
$M_{\rm l} = 5\times 10^{7}{\rm M}_{\odot}$. 
Fig.\ \ref{nfig:1}, for instance, shows
the average number of progenitors in each tree as a function
of mass, at four different redshift steps. The theoretical spectrum can be 
calculated analytically from EPS theory, as 
\begin{eqnarray}
n_{\rm p}(z,M_1){\rm d}M_1 = {{M_0}\over{M_1}}f_{S_1}(S_1,\omega_1|S_0,\omega_0){\left|{{{\rm d}S}\over{{\rm d}M}}\right|}{\rm d}M_1\,,
\end{eqnarray}
where $M_0$ is the mass of the main halo, $S_0 = \sigma^2(M_0)$,
$S_1 = \sigma^2(M_1)$, $\omega_0$ is the critical overdensity 
at the present day, $\omega_1$ is the critical overdensity 
extrapolated back to $z$, 
and $f_{S_1}$ is the first merger probability given in LC93 
(equation 2.15). 
The solid lines show this prediction. The merger trees
reproduce the expected distribution extremely well over 4 orders of 
magnitude in mass.
The only slight discrepancy is at very low redshift,
when the finite redshift step size reduces the number of low-mass
haloes somewhat.

\begin{figure}
  \centerline{\psfig{figure=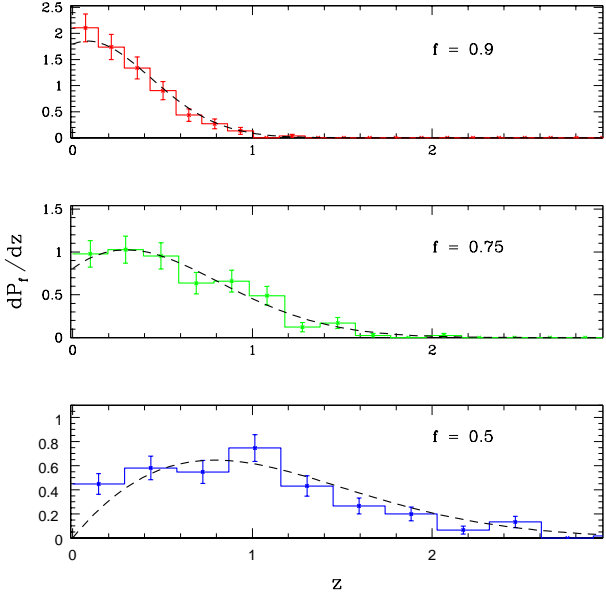,width=1.0\linewidth,clip=,angle=0}}
  \caption[]{The distribution of formation
redshifts for SCDM trees, compared to the analytic prediction, for
$f = 0.9$, $f = 0.75$ and $f = 0.5$ 
(top, middle and bottom panels respectively).}
\label{nfig:2}
\end{figure}

As a simple test of the global properties of our merger trees, we 
can calculate the formation epoch of the main halo in each tree,
that is the time when the main system had assembled some fraction
$f$ of its final mass. The resulting distribution can be
calculated analytically in PS theory:
\begin{eqnarray}
P_{f}(z,M_0) = \int^{S_{f}}_{S_0}{{M_0}\over{fM_0}}f_{S_1}(S_1,\omega_1|S_0,\omega_0){\rm d}S_1\,,
\end{eqnarray}
(LC93, equation 2.26), where $S_{f} = \sigma^2(fM_0)$ and
the other variables are as above. 
Fig.\ \ref{nfig:2} shows the distribution of formation
redshifts for the SCDM trees used in this paper, compared to the analytic 
distribution, for $f = 0.9$\,, $f = 0.75$ and $f = 0.5$ (denoted $z_{f,90}$, 
$z_{f,75}$ and $z_{f,50}$ respectively).
There is good agreement with the analytic estimate
down to $f = 0.5$, where the definition of $z_f$ breaks down as
a present-day halo of mass $M$ may have more than one progenitor with
a mass of $fM$. 

\begin{figure}
  \centerline{\psfig{figure=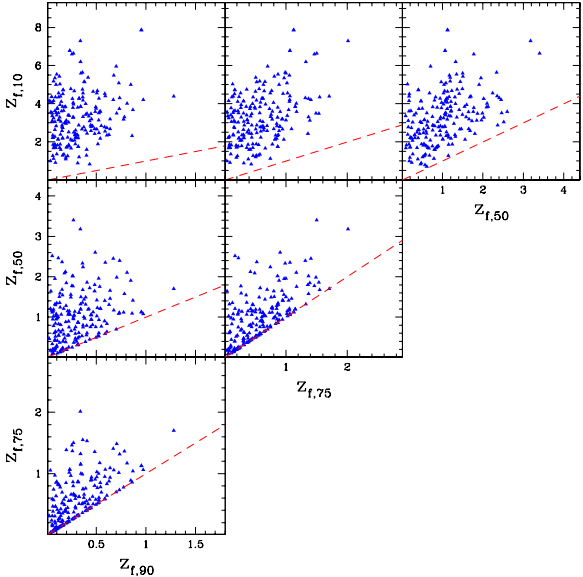,width=1.0\linewidth,clip=,angle=0}}
  \caption[]{Correlations between the formation epochs defined using
different values of $f$.
The dashed line shows the region excluded by the requirement that the halo
grow monotonically.}
\label{nfig:3}
\end{figure}

We will show in paper II that many global properties of
substructure correlate strongly with the relative age of the
halo. It is not clear what value of $f$ to choose when estimating
the age of a system. In
Fig.\ \ref{nfig:3} we show a comparison between the formation
epochs defined for different values of $f$, for individual merger trees. 
Beyond the limits imposed by the requirement that haloes grow 
monotonically, i.e. that $z_{f,90} < z_{f,75} < z_{f,50}$ 
(dashed lines), 
there is only a slight correlation between the different formation times,
with fairly large scatter. We will show in part II that one can distinguish
different effects in the spectrum of halo substructure, using the different 
formation epochs to characterise separate phases in the mass accretion 
history of a given halo.

\subsection{Structural properties of the main system}\label{subsec:2.2}

The spherical collapse model described in section \ref{subsec:2.1} 
also provides an estimate of the final density of the virialised region 
of a halo. If one assumes that a collapsing density fluctuation virialises 
through violent relaxation 
when the formal solution reaches zero radius, and that virialisation 
conserves the initial energy of the system, then virialisation should occur 
when the mean density of the system relative to the critical density reaches 
a specific value $\Delta_{\rm c} \equiv \rho_{\rm vir}/\rho_{\rm c}$. 
The precise value of $\Delta_{\rm c}$ has been calculated for various
CDM cosmologies (e.g.\ Peebles 1980 or Padmanabhan 1993 for SCDM; 
Maoz 1990 or LC93 for OCDM; Kochanek 1995 or 
Eke, Cole \& Frenk 1996 for LCDM) and lies in the range 100--200.
Although this analytic estimate ignores several complications in the
evolution of overdense regions, simulations show that it is approximately
correct (Lacey \& Cole 1994; Navarro, Frenk and White 1997, 1997, 
NFW96 and NFW97 hereafter). 

In what follows, we will therefore assume that the virial density is
exactly $\Delta_{\rm c}\rho_{\rm c}(z)$ at any epoch $z$, so the radius 
of a halo of mass $M$ at that epoch is simply 
\begin{equation}
r_{\rm vir}(z) = \left({{3\,M(z)}\over{4\pi\Delta_{\rm c}\rho_{\rm c}(z)}}\right)^{1/3}\,,
\end{equation}
and circular velocity at the virial radius is:
\begin{equation}
V_{\rm c, vir}(z) = \left({{G\,M(z)}\over{r_{\rm vir}(z)}}\right)^{1/2}\,.
\end{equation} 
We note that in the EPS picture, all haloes seen at a given epoch have 
{\it the same mean density} within their virial radius, independent 
of mass. We do expect systematic differences in the density distribution 
within the virial radius for haloes of different masses, however, as 
discussed below.

\subsubsection{The universal density profile}

The spherical collapse model does not specify how dark matter
behaves interior to the virial radius. One of the most
important results of numerical simulations of halo formation was 
the discovery that the spherically averaged density profile of haloes 
has a characteristic form which is independent of cosmology 
(NFW96; NFW97; Moore et al.\ 1998). 
In its outer regions, close to the virial radius, this universal density 
profile has a logarithmic slope of $-3$; the slope then decreases to $-2$ 
at a characteristic radius near the peak of the rotation curve, and then drops 
further to a central slope of less than $-2$.

Two analytic forms commonly used to describe the simulation results
are the Navarro, Frenk and White (NFW) profile (NFW96; NFW97):
\begin{equation}
\rho(r) = {{\rho_{\rm s}\,r_{\rm s}^3}\over{r\,(r + r_{\rm s})^2}}
\end{equation}
and the Moore profile (Moore et al.\ 1998):
\begin{equation}
\rho (r) = {{\rho_{\rm s}\,r_{\rm s}^3}\over{r^{1.5}\,(r^{1.5} + r_{\rm s}^{1.5})}}
\end{equation}
In their outer regions, these profiles are almost identical if
one normalises the scale radius $r_{\rm s}$ and density $\rho_{\rm s}$
such that they match at the radius where the circular velocity peaks, 
$r_{\rm p} = 2.163\,r_{\rm s}({\rm NFW}) = 1.25\,r_{\rm s}({\rm M})$. 
A similar, non-analytic profile with a central inner slope of $-0.75$ 
was described by Taylor \& Navarro (2001); 
it has a scale radius of 
$r_{\rm s}({\rm TN}) = 1.67\,r_{\rm s}({\rm NFW})$, 
and also matches the universal
profile in its outer regions if normalised as described above. 
The precise value of the inner logarithmic slope is still 
controversial, however, as discussed in the next section. 

\subsubsection{The central cusp}

The slope of the central cusp in haloes is particularly hard to determine
from simulations, as it is strongly affected by softening, mass and force
resolution. Initially, simulations indicated the central cusp had a 
logarithmic slope of $-1$ (NFW96, NFW97). Subsequent
convergence tests at higher resolution suggested that the slope
was steeper, possibly as steep as $-1.5$ (e.g.\ Moore et al.\ 1998; 
Fukushige \& Makino 2001). The most recent high-resolution
simulations find that while the central slope is less than $-1$, there 
may be more mass in the central regions of haloes than predicted by the NFW 
formula (Power et al.\ 2003). A further complication is that the logarithmic
slope may continue to change slowly, not approaching its final value
until it is well below the resolution limit of current simulations 
(Taylor \& Navarro 2001). When considering subhalo evolution, the amount 
of mass 
in the central regions is generally more important than the exact slope
of the density profile, so we will use the analytic Moore profile in our model
by default, to account for the extra mass seen by Power et al.\ (2003). 
We expect results for the TN profile to be similar, as explained in 
section \ref{subsec:disruption}. In paper II we will consider 
the effect of other density profiles, such as an NFW profile, on our results.

\subsubsection{Halo concentration}\label{subsubsec:2.2.3}

The one remaining parameter required to specify a halo's density
profile completely is its concentration, defined as 
$c \equiv r_{\rm vir}/r_{\rm s}$, 
which measures the position of the break radius relative to the virial radius 
of the system. Generally low-mass haloes, which form at early times, are more 
concentrated (i.e.\ $r_{\rm s}$ is smaller relative to $r_{\rm vir}$) than 
massive haloes, which form later. It was first suggested that the break in 
the universal density profile marks the boundary between material built up 
during the initial formation of the halo or its precursors at early times, 
and material accreted on to these cores at later times (NFW96, NFW97). 
There is now some direct confirmation 
of this picture from simulations (Wechsler et al.\ 2002; Zhao et al.\ 2002), 
and there are several 
predictions of the concentration of haloes as a function of their mass and the 
redshift, based on this interpretation (NFW96; Bullock et al.\ 2001b; 
Eke Navarro and Steinmetz 2001 -- ENS01 hereafter). 
We will assume the concentrations predicted by the ENS01 
model, the most recent fully-analytic model. We calculate ENS01 
concentrations using code made publicly available by the authors, 
and convert these to Moore concentrations using the ratio 
of scale radii given above, $c_{\rm M} = (1.25/2.163)\,c_{\rm NFW}$.
We note that for very massive haloes or at very high redshifts, 
the ENS code predicts concentrations of less than 1, presumably
because these objects are more massive than the typical mass assumed
to have collapsed by that redshift. Since the authors do not
measure any concentrations below $\simeq$ 1--2 in their analysis of the
simulations, we
assume $c_{\rm NFW} = 1$ is the minimum concentration for a bound halo 
and ignore concentrations below this. The exact value of the minimum
should be unimportant in practice, since almost no haloes in our merger
trees have concentrations this low.

 We also note that there is both a large scatter in measured halo 
concentrations 
relative to the ENS01 predictions or those of other analytic models based 
on average 
properties, and that this scatter correlates with the mass accretion 
histories. 
The more recent models of Wechsler et al.\ (2002) and Zhao et al.\ (2002), 
may be more accurate in this respect, but they require a detailed 
knowledge of the mass accretion history, which will not always be available 
for all the systems in our model. Our assumed halo concentrations may have some
effect on our results, particularly the concentration of the central potential
in the main system. We will discuss this issue further in paper II.

\subsection{Mergers}\label{subsec:2.3}

For an individual halo, the mass $M(z)$ can be calculated as described in
section \ref{subsec:2.1}.
In the EPS approach, sudden jumps in this mass accretion history are 
interpreted as merger 
events, where one or more distinct virialised haloes join up 
with the first system. Following these mergers
back in time then produces full merger `trees', as discussed previously. 
The resulting merger histories contain no spatial 
information about merger events, however. In order to relate individual 
merger events to infalling subhaloes, we need to examine the EPS 
definition of a merger
in more detail.

\subsubsection{Timing and orbital energy}

In EPS models, two haloes `merge' when they are both inside a region
with a characteristic density $\Delta_{\rm c}\rho_{\rm c}(z)$. This 
same density characterises the virialised region of 
an isolated halo. Thus when a small halo of mass $M_1$ and virial radius 
$r_1$ merges with a halo of mass $M_2 >> M_1$, the virial radius of the new, 
combined system will be $r_{1+2} \simeq r_2 >> r_1$, and thus 
the actual `merger', that is the moment when the volume containing
both haloes reaches the density $\Delta_{\rm c}\rho_{\rm c}(z)$, 
will occur roughly 
when the smaller halo crosses the virial radius of the larger system 
for the first time. More generally, even if $M_1$ 
is larger relative to $M_2$, the merger will still occur roughly when 
the infalling halo first crosses a spherical boundary of radius 
$r_{1+2} = (3(M_1+M_2)/4\pi\Delta_{\rm c}\rho_{\rm c}(z))^{1/3}$ 
around the main halo. 
We will take this precise moment as the event recorded in the merger
tree. Once this first infall has occurred, we include the mass of the 
satellite as part of the main system when
calculating its potential, since this is the mass assumed in the merger 
tree at subsequent times. 
(This introduces a slight inconsistency in our orbital calculations
-- satellites move in a potential that includes their own mass --
but the effect should be minor except for major mergers, which evolve
very quickly, and where we do not expect our simplified orbital calculations 
to be accurate in any case.)

What velocity should one assign to a merging halo? 
In the spherical collapse model, assuming no shell-crossing (that is 
assuming the merging halo falls in radially under the influence of a 
constant mass interior to its radius, and experiences no other forces 
beyond this), then the velocity of the subhalo when it `merges' will be 
\begin{equation}
V_{\rm infall} = V_{\rm c,vir} = \left({{GM_{\rm vir}}\over{r_{\rm vir}}}\right)^{1/2}\,, 
\end{equation}
where $M_{\rm vir}$ is the mass of the whole system, including the new
subhalo, within its virial radius $r_{\rm vir}$. 
Numerical simulations confirm this prediction;
Tormen (1997), for instance, finds an average velocity for merging satellites
(in the sense defined above) of $1.1 \pm 0.1\ V_{\rm c, vir}$.
In what follows we will assume that $V_{\rm infall} = V_{\rm c,vir}$
exactly, for single merger events. (Infalling groups will have an additional
scatter in their velocities, as explained in section \ref{subsec:5.3}.)

\subsubsection{Angular momentum}

The angular momentum of infalling satellites is harder to estimate, 
although one can attempt to derive it by tidal torquing arguments 
(Peebles 1969), or by assuming a density distribution and a background 
potential (van den Bosch et al.\ 1999).
A convenient parameterisation for this quantity is the circularity
of the orbit $\epsilon \equiv L/L_{\rm c}$, the ratio of the (initial) angular 
momentum $L$ to the angular momentum of a circular orbit with the same energy,
$L_{\rm c}$ (LC93).  
Several high-resolution simulations of individual haloes suggest that 
$\epsilon$ has a roughly Gaussian distribution between 0 and 1, with 
a peak at 0.5--0.55 and a dispersion of 0.2--0.3 
(Navarro, Frenk \& White 1995; 
Tormen 1997; Ghigna et al.\ 1998). We note that for major mergers, 
the lower-resolution but 
larger-volume simulations of Vitvitska et al.\ (2002) find larger values of 
$V_{\rm infall}/V_{\rm c,vir}$ and a more skewed distribution of angular 
momenta. This is partly because they normalise to the circular velocity 
of the most massive halo in a pair, however, and this may be substantially 
smaller than the final circular velocity of the merger remnant in the case 
of a major merger. Since they only consider a few outputs of their 
simulation, it is 
also not clear whether the satellites have been caught on their first infall 
into the larger system. Thus we will assume the high-resolution results are 
more indicative of the initial angular momentum distributions of satellites.

In section \ref{sec:static}, we will consider satellite evolution for various 
specific values of $\epsilon$. In section \ref{sec:dynamic} and in our full 
model, we will use a Gaussian distribution with a mean 
${\overline \epsilon} = 0.4$ and dispersion $\sigma_{\epsilon} = 0.26$,
which after selective disruption of the more radial orbits gives
a mean of ${\overline \epsilon} = 0.55$ and a variance of 
$\sigma_{\epsilon} = 0.23$ in the surviving satellites, similar to
the distribution in Tormen (1997) or Ghigna et al.\ (1998). 
In paper II we will also discuss
the effect of changing the energy and angular momentum distributions.

\subsubsection{Structure of the merging halo}

In the EPS model, as noted above, distinct haloes merging at any 
epoch will have the same average density within their virial radius, 
independent of mass. 
Furthermore, they should also obey the concentration relations 
described in section \ref{subsubsec:2.2.3}. 
Thus, given the epoch of a merger and the mass of 
the infalling halo from a merger tree, we can also calculate its
initial virial radius, scale radius and density profile. This
is enough to describe the infalling system completely, neglecting 
higher-order substructure it may contain. We will consider this 
final complication next.

\subsection{Substructure and self-similarity}

From the discussion in the previous section, 
within our simplified description of structure formation the 
initial conditions for cosmological mergers between haloes are well defined. 
Specifically, an individual event within a merger tree corresponds to a halo 
with a well-defined density profile, on an orbit chosen from a specific 
distribution, crossing the virial radius of the new, combined
system for the first time. 

As a halo merges, it will be stripped of its outer regions. 
The remaining core will be denser than 
most of the main system it orbits in, and may persist as a distinct subhalo 
for many orbits. Since the infalling system has itself formed 
through hierarchical merging, however, 
it should contain its own dense substructure. This `higher-order'
substructure may survive within the main system even after its original
parent has been disrupted, just as galaxies merging
into a cluster as part of a small group may be stripped from the group,
but will continue to survive as distinct objects. Much of this hierarchy
will not be resolved in even the largest of current simulations, but it
may make an important contribution to halo substructure, as shown in
section \ref{sec:pruning}.

This leads to a tricky problem when modelling substructure in 
hierarchically
assembled haloes. The statistical properties of substructure can be
determined most accurately by considering the evolution of individual 
subhaloes in a simplified system. A full description of substructure should 
really be applied recursively, however, considering the effect of substructure 
within substructure and so one. Thus the process of developing and testing
a model of substructure is necessarily iterative. 
The model used to include substructure in a merger tree should also be
computationally efficient, treating each step of the merging process with
only a few calculations, so that it will be able to handle many levels of 
the merger hierarchy. 

To begin with, in the next section we will consider the evolution
of subhaloes in a very simple system, in order to establish certain basic
patterns. While the density profiles and orbits will be chosen as described 
above, we will fix the mass, concentration and density of the main 
system, and give the infalling subhaloes the same density as the main
system. Thus,
in effect we will be considering the evolution of a system cutoff from 
hierarchical
growth at some point, and left to evolve without any further accretion
or change in its structural properties.
In section \ref{sec:dynamic}, we will 
then consider the evolution of subhaloes
in evolving systems, with model parameters that have been chosen after 
iterating several times. 
Finally, in section \ref{sec:pruning} we will explain how to include 
this information in merger trees, and how
we have adjusted the parameters in our model by iteration.

\section{Evolution of substructure in a static system}\label{sec:static}

\subsection{Review of previous work}

In TB01, we 
developed a semi-analytic model for the dynamical evolution of
spherical satellites in the potential of a larger system. This
model includes simple descriptions of the most important physical
processes that determine satellite evolution, 
namely dynamical friction, tidal stripping, and tidal heating.
Dynamical friction, the drag force produced as a massive object
moves through a background of particles, is modelled using Chandrasekhar's
formula (Chandrasekhar 1943), with the Coulomb logarithms left as free
parameters. Tidal stripping is modelled by assuming that material
outside the instantaneous tidal radius of the satellite is stripped
off on a time-scale equal to the dynamical time of the satellite
at its half-mass radius. By introducing this time-scaling, we were
able to capture the dependence of mass loss on orbital pericentre
and apocentre, which is not reproduced in simpler treatments (Taylor 2001). 
Finally, to model the rapid shocks experience by the satellite when it
passes through the pericentre of its orbit or through the plane of 
a galactic disk,
we include a heating term in the expression for mass loss, with
an adjustable heating coefficient. 

Overall, the model in TB01 thus predicts orbital evolution and mass 
loss for satellites using three free parameters; a Coulomb logarithm 
$\Lambda_{\rm s}$ for the halo and other spherically distributed
material, a Coulomb logarithm $\Lambda_{\rm d}$ for the galactic disk, 
and a heating coefficient $\epsilon_{\rm h}$.
Comparing model predictions to a set of 15 high-resolution simulations 
of encounters between satellites and disk galaxies by 
Vel\'{a}zquez and White (1999), we found we were able to reproduce the orbital 
evolution and mass-loss history of the
satellites to within 10--20 per cent over most of their evolution (until they
had lost $\simeq 90$ per cent of their mass), using as single set of 
parameters,
 $\Lambda_{\rm s} = 2.4$, $\Lambda_{\rm d} = 0.5$ and 
$\epsilon_{\rm h} = 3.0$. The latter value of the heating coefficient
also produces quite a good fit to the mass-loss rates
measured in a set of high-resolution simulations by Hayashi \& Navarro 
(Hayashi et al.\ in preparation; see Taylor 2001 for a comparison), 
which use
very different potentials, orbits and density profiles, 
so we have reason to believe it is generally applicable.

TB01 did not discuss the internal structural evolution of satellites.  
In subsequent work by Hayashi et al.\ (2003), the evolution of the density
profile was found to depend on the total mass loss alone, rather than
on the details of the mass-loss history. This paper provides formulae
for determining the density profile and characteristic radii of a
satellite at any point in its evolution, as well as an
estimate of when repeated mass loss will disrupt an object completely.

Taken together, these results constitute a full analytic description
of the dynamical evolution of satellites in realistic potentials,
and one that is easily applicable to a large number of objects with 
minimal computational effort, as uses only the global properties
of satellites, such as their total mass and the characteristic radii 
of their density profile. 
We will now apply this analytic description to the problem of subhalo 
evolution. We take the same values for the dynamical parameters,
$\Lambda_{\rm s} = 2.4$ and $\epsilon_{\rm h} = 3.0$, determined in TB01,
and scale $\Lambda$ by $(M_{\rm sat}/M_{\rm halo})^{-1}$  (where
$M_{\rm sat}$ is the mass of the satellite and $M_{\rm halo}$ is the
mass of the main system), as discussed in section 4.1 of TB01.
There is no need to specify $\Lambda_{\rm d}$, since the potentials
considered in this paper have no disk component.

\subsection{The orbital time-scale}\label{subsec:3.2}

First, we will examine the basic properties of the orbits specified
in section \ref{subsec:2.3}, that is orbits starting at the virial 
radius of a large halo
with a Moore density profile, with a total initial velocity of
magnitude $V_{\rm c}$ and various possible circularities. We choose
a concentration of $c_{\rm M} = 10$ for the main system, the value 
predicted by the 
ENS01 concentration relations for a halo similar to that of the Milky Way,
at $z = 0$ in our SCDM cosmology. Our fiducial system has a mass 
of $1.6 \times 10^{12} {\rm M}_{\odot}$ and a virial radius of 314 kpc, 
but  
our results scale in a straightforward way and can all be described in terms
of scaled variables. Given this scaling, the evolution of a satellite 
subhalo in our model will depend only on its mass relative to the mass of 
the main system, its concentration, the initial circularity of its orbit, 
and the concentration of the main system. In most of this section we
will also ignore dynamical friction, to simplify the analysis of the dynamics.
Thus for a given main potential, the evolution of a satellite will depend only 
on its initial concentration and its circularity.

\begin{figure}
  \centerline{\psfig{figure=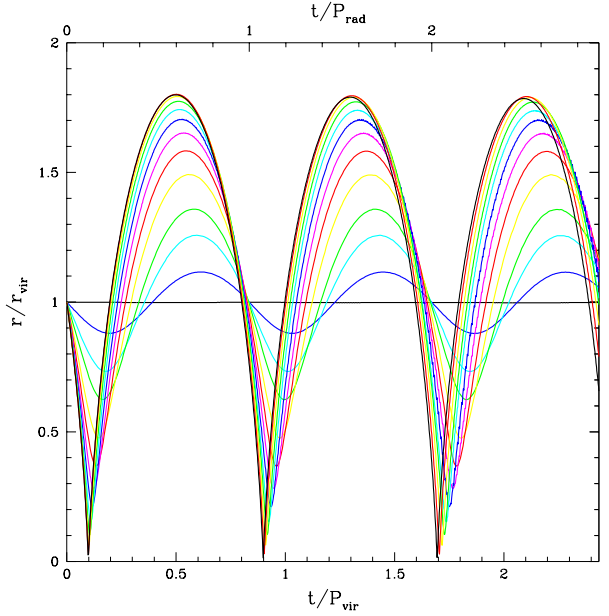,width=1.0\linewidth,clip=,angle=0}}
  \caption[]{Radius (in units of the virial radius) versus time (in units
of the azimuthal period of a circular orbit at the virial radius,
$P_{\rm vir} = 2\pi\,r_{\rm vir}/V_{\rm c, vir}$\,), for orbits of circularity
$\epsilon =$\ 1.0, 0.99, 0.95, 0.9, 0.8, 0.7, 0.6, 0.5, 0.4, 0.3, 0.2, 0.1, 
and 0.05.
The top axis shows time in units of the period for small radial oscillations
$P_{\rm rad} = 2\pi/\kappa$ (see text). Dynamical friction has not 
been included.}
\label{nfig:4}
\end{figure}

We will see in the section \ref{subsec:3.3} that mass loss occurs primarily at
the pericentric passages in a subhalo's orbit. Thus the characteristic 
time-scale for the evolution of substructure will be the period of
radial oscillations (or `radial period'). 
Fig.\ \ref{nfig:4} 
shows the radial coordinate of satellites on orbits of various 
different circularities, ranging from $\epsilon =$\ 0.05 to 
$\epsilon =$\ 1.0 (different curves), 
with the vertical axis scaled by the virial radius of the main system and 
the horizontal axis scaled by the (azimuthal) period of a circular orbit 
at the virial radius, $P_{\rm vir} = 2\pi r_{\rm vir}/V_{\rm c, vir}$. 
(The latter quantity is equal to $H_0^{-1}\,(8\pi^2/\Delta_{\rm c})^{1/2}$ 
from the spherical collapse model; for the SCDM cosmology considered here
$\Delta_{\rm c} = 18\pi^2$, and thus $P_{\rm vir} = 2/3\,H_0^{-1}$ 
is simply the age of the universe at the epoch which characterises the
virial density of the main system.)
Dynamical friction has been ignored, as mentioned earlier, 
so the orbits shown in Fig.\ \ref{nfig:4} correspond to those 
of low-mass satellites. 

\begin{figure}
  \centerline{\psfig{figure=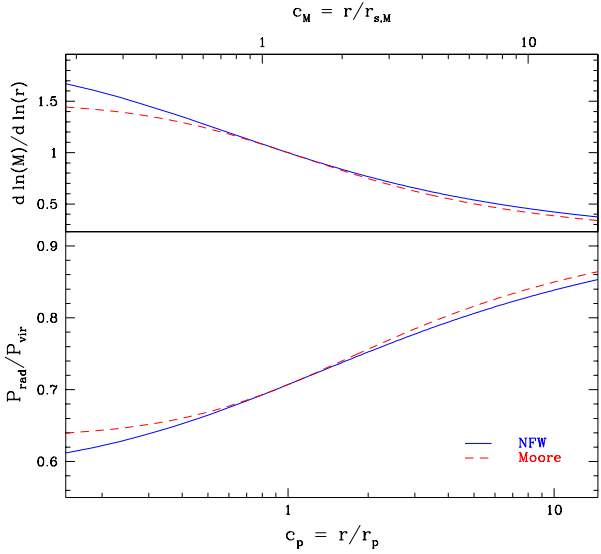,width=1.0\linewidth,clip=,angle=0}}
  \caption[]{The logarithmic derivative of the mass profile (top panel) 
and the radial period at the virial radius $P_{\rm rad}$ (bottom panel) 
versus concentration, for NFW and Moore profiles (solid and dashed lines
respectively).}
\label{nfig:5}
\end{figure}

The variation in radius is regular and periodic, but clearly the period 
differs from $P_{\rm vir}$. In fact, it is roughly equal to the period 
for small radial oscillations at the virial radius, 
$P_{\rm rad} \equiv 2\pi/\kappa$, where $\kappa$ is the epicyclic 
frequency:
\begin{equation}
\kappa = {{V_{\rm c}}\over{r_{\rm vir}}}\left(1 + {{{\rm d}\ln M}\over{{\rm d}\ln r}}\right)^{1/2}\,,
\end{equation}
as expected from orbital theory for a spherical system with mass $M$ 
interior to $r$ (Binney and Tremaine 1987, section 3.2.3).
This quantity will vary with the slope of the density profile. 
Fig.\ \ref{nfig:5} 
shows the logarithmic slope (top panel) and the ratio of the radial and
azimuthal periods (bottom panel) at the virial radius as a function of the 
concentration of 
the system, for haloes with NFW (solid) or Moore (dashed) profiles.
For Moore profiles with concentrations typical of galaxy haloes, 
$P_{\rm rad} \simeq 0.835 \pm 0.015\, P_{\rm vir} \simeq 5.24(r_{\rm vir}/V_{\rm c, vir})$. 
Strictly speaking, this radial period should apply only to small amplitude 
oscillations about a circular orbit, but from Fig.\ \ref{nfig:4} 
we can see that it 
describes the period of radial oscillations reasonably well even for 
extremely non-circular orbits. There is a slight change in the radial period
with orbital circularity -- the oscillations in radial orbits 
are about 5 per cent faster than those in almost circular orbits -- but 
the main difference is a shift of the time of pericentric passage within 
an orbit from 0.125\,$P_{\rm rad}$ to 0.25\,$P_{\rm rad}$
as $\epsilon$ increases. In what follows, we will assume $P_{\rm rad}$ 
is the basic period for subhalo orbits, and that pericentric passages
occur somewhere between 1/8 and 1/4 of a radial period, and then
periodically thereafter. 

Finally, although we have calculated orbits in the potential generated 
by a Moore density profile, we note that all but the most
radial orbits would be very similar in an NFW profile (of if we had
added a small galactic component to the potential) since most orbits
do not enter into the central regions of the potential where the profiles 
differ (e.g.\ in Fig.\ \ref{nfig:4}, only orbits with $\epsilon < 0.2$
reach radii of less than $0.1\,r_{\rm vir}$).
We also note that the extremely radial orbits plotted in this section 
require very small time-steps to integrate properly. For realistic 
distributions of orbital circularity, few subhalo orbits are this 
expected to be this radial, as discussed below. Since the mass-loss
model developed in TB01 is unlikely to be accurate for orbits with very
small pericentres in any case, in our full model we will consider subhaloes
disrupted when they come very close to the centre of the 
main system ($r < 0.01\,r_{\rm vir}$), and stop following their orbital 
evolution at that point. Otherwise, subhalo 
orbits are calculated with an adaptive time-step, sufficiently small to 
provide reasonable accuracy given this cutoff at very small radii.

\subsection{Mass loss}\label{subsec:3.3}

We proceed to consider mass loss on typical orbits. From the
previous discussion, the mean density of haloes (within their virial radius)
when they merge will be the same as the mean density of the main system, 
and their initial mass will be specified by the merger tree. 
If we fix the density profile and concentration 
of the main system, assume the subhalo has a Moore density profile, 
and ignore dynamical friction, then the evolution of a subhalo will depend 
only on its concentration and on the circularity of its orbit.
In this section we will study the case 
$c_{\rm M}$\,(sat) $= c_{\rm M}$\,(main) $= 10$; we will
then show results for a typical range of concentrations below. We
include the effects of tidal shocks, as in TB01, but choose parameters 
appropriate
to slow shocks (an adiabatic coefficient $\gamma = 1.5$ and a shock criterion
$t_{\rm shock} < 4\,t_{\rm orb, sat}$; see Gnedin \& Ostriker (1999) or 
TB01 for an explanation of these 
parameters), since the passage through pericentre is slower than
the passage through a disk considered in TB01.

\begin{figure}
  \centerline{\psfig{figure=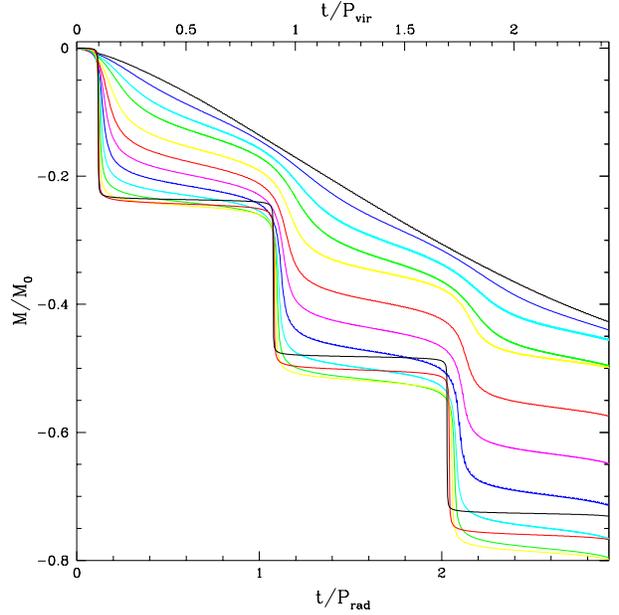,width=1.0\linewidth,clip=,angle=0}}
  \caption[]{Bound mass fraction versus time, in units of the radial period 
$P_{\rm rad}$ (bottom axis) or the azimuthal period $P_{\rm vir}$ 
(top axis). The curves are for the same circularities as in Fig.\ \ref{nfig:4},
and dynamical friction has not been included, as before.}
\label{nfig:6}
\end{figure}
 
Fig.\ \ref{nfig:6} shows the fraction of the subhalo's original mass that 
remains bound as a function of time, for the same set of orbits plotted 
in Fig.\ \ref{nfig:4}. We have estimated the bound mass using the
model from TB01, which assumes that a fraction $\Delta t/t_{\rm d,h}$ 
of the mass outside the instantaneous tidal radius is lost in a time step 
of length $\Delta t$ (where $t_{\rm d,h}$ is the dynamical time of the 
system at its half-mass radius), and also includes a heating term to correct 
for rapid tidal shocking. This model
reproduces the pattern of continuous mass loss throughout the orbit, with
a sharp increase at each pericentric passage, that has been seen in
simulations of tidal stripping and heating. For the most radial orbits,
almost all the mass loss occurs at pericentre, while for circular orbits
mass loss is slower but continuous. In general, satellites on
more radial orbits lose more mass, mainly because they pass through 
smaller pericentres and experience stronger tidal fields. This trend
is not completely monotonic, however, since a satellite on an extremely
radial orbit spends very little time close to the pericentre of its orbit,
where tides are strongest. Comparing the most extreme curve in 
Fig.\ \ref{nfig:6} to slightly less radial orbits, we see that while the 
mass loss at the first pericentric passage is similar, 
less mass is lost immediately afterwards, as the satellite moves
away from the pericentre of its orbit faster than it would if its
orbit were more circular.

\begin{figure}
  \centerline{\psfig{figure=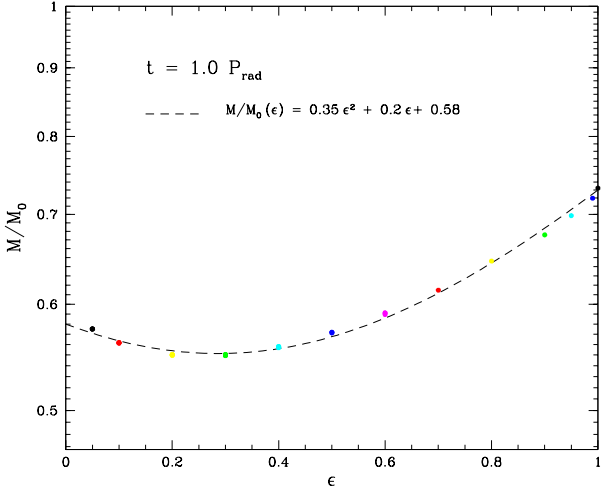,width=1.0\linewidth,clip=,angle=0}}
  \caption[]{Bound mass fraction after one radial period 
($t = 1.0\,P_{\rm rad} \simeq 5.24\,r_{\rm vir}/V_{\rm c}$),
as a function of circularity $\epsilon$, for subhaloes with a 
Moore density profile of concentration $c_{\rm M} = 10$ and no dynamical 
friction. The dashed curve shows the
functional fit $M/M_0 = 0.35\,\epsilon^2 + 0.2\,\epsilon + 0.58$.}
\label{nfig:7}
\end{figure}

Fig.\ \ref{nfig:7} shows the fraction of the original mass which remains 
bound after one radial period, as a function of orbital
circularity. The general trend towards more mass loss on radial 
orbits is clear, but orbits with $\epsilon \simeq 0.2$--0.3 
produce the most net mass loss per radial period.
A rough fit to the bound mass fraction after one
radial period as a function of circularity,
$M/M_0 \simeq 0.35\,\epsilon^2 + 0.2\,\epsilon + 0.58$, is also shown.

The scaling of mass loss with orbital and halo properties is particularly
important for haloes with cosmological profiles, that is profiles with
a central region that is sub-isothermal. Within some radius, the total 
energy of these systems is actually positive, as there is insufficient 
mass to bind the material in the core (Hayashi et al.\ 2003). 
Thus repeated mass loss may disrupt such systems completely 
if they cannot readjust themselves into a new virial equilibrium quickly 
enough. We discuss this point in more detail in the next section. 

\subsection{Disruption}\label{subsec:disruption}

\subsubsection{The disruption criterion}
One important question regarding the evolution of substructure is whether it 
can ever be fully disrupted by tidal forces, and if so when this occurs. 
This point is very hard to resolve using N-body simulations alone, 
since subhaloes become more and more sensitive to purely numerical effects
as they lose mass.
One basic analytic approach to determining the stability of a system
is to calculate its net kinetic and potential energy interior to some
radius (calculated as if the system were truncated at that radius), 
prior to any heating or stripping:
\begin{eqnarray}
E(<r) &=& K(<r) + W(<r) \nonumber\\
&=& \int_0^r\rho (r')\,\overline{v^2}(r')\,2\pi\,r'^2{\rm d}r' \nonumber\\
&+&  \int_0^r\rho (r')\,\phi (r',r)\,2\pi\,r'^2{\rm d}r'\,,
\label{eq:disrupt}
\end{eqnarray}
where $\overline{v^2}(r')$ is the velocity dispersion at $r'$ and 
$\phi (r',r)$ is the potential at $r'$ generated by material within $r$.
If tidal stripping removes material from the outer part of a system without
affecting the distribution function of the remaining material too
much, then the system should be unstable once it is stripped to
a radius $r_{\rm bind}$ within which the total energy, as defined above, is 
positive.  

For a given density profile, we can calculate the initial critical radius
using equation (\ref{eq:disrupt}). 
For an NFW profile, $r_{\rm bind} = 0.77\,r_{\rm s}\,({\rm or}\ 0.353\,r_{\rm p})$, 
for a TN profile    $r_{\rm bind} = 0.399\,r_{\rm s}\,({\rm or}\ 0.307\,r_{\rm p})$,
and for a Moore profile $r_{\rm bind} = 0.37\,r_{\rm s}\,({\rm or}\ 0.296\,r_{\rm p})$, 
while for an 
isothermal profile the critical radius is infinite, since the total energy,
as defined above, is negative at all radii.

This stability criterion has the important consequence that systems
with sub-isothermal cusps or cores will become unstable and disrupt 
rapidly once they have lost a large fraction of their mass. 
In fact, in the most recent and most detailed numerical study 
of mass loss (Hayashi et al.\ 2003), it was found that systems on
circular orbits disrupt
if their tidal radius $r_{\rm t}$ is less than $2\,r_{\rm bind}$.
Unfortunately, for general orbits the criterion is not straightforward 
to apply,
since $r_{\rm t}$ varies with time and $r_{\rm bind}$ also varies 
as the density profile and distribution
function change due to mass loss. If we assume that a system re-virialises
instantaneously whenever it loses mass, for instance, then 
the scaling formula for the stripped
density profile proposed in Hayashi et al.\ (2003) (equations 8--10) 
predicts that a halo with an NFW profile of concentration $c_{\rm NFW} = 10$ 
will never be disrupted by mass loss, as its total mass will
always exceed the mass inside $2\,r_{\rm bind}$. Nonetheless, these
formulae do predict a point of minimum stability (in the sense that the
tidal radius of the system is close to $2\,r_{\rm bind}$) when the system
has lost approximately 97 per cent of its original mass, and $r_{\rm bind}$
is half the original critical radius $r_{\rm bind, 0}$. Empirically, 
Hayashi et al.\ trace the evolution of systems on radial orbits down
to about this level, while for circular orbits they show some systems
still bound when they have lost all but $0.3$  per cent or more of their mass,
corresponding to a critical radius of roughly $0.1$ of the original.

\begin{figure}
  \centerline{\psfig{figure=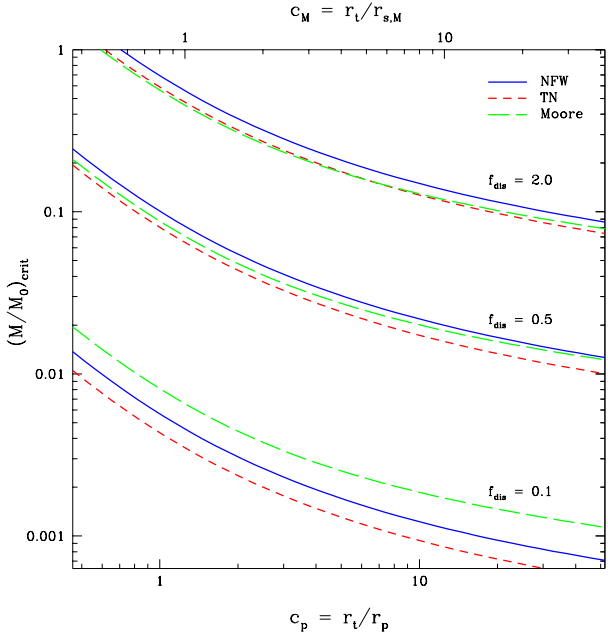,width=1.0\linewidth,clip=,angle=0}}
  \caption[]{The mass below which a system of a given concentration will be 
disrupted, as a fraction of its original mass, for disruption parameters 
$f_{\rm dis} = 2.0$, $f_{\rm dis} = 0.5$, and $f_{\rm dis} = 0.1$. The solid 
line is for an NFW profile, the short-dashed line is for a TN profile and 
the long-dashed line is for a Moore profile. The concentration is expressed 
in terms of the peak radius on the bottom axis, or the equivalent Moore
concentration on the top axis.}
\label{nfig:8}
\end{figure}

In what follows, we will parameterise the uncertainty in the disruption
criterion by assuming that systems disrupt when 
$r_{\rm t} < f_{\rm dis}\,r_{\rm bind, 0}$, or $M < M_{\rm bind} \equiv
M(<f_{\rm dis}\,r_{\rm bind, 0})$. 
Motivated by the simulations, we will consider results for 
$f_{\rm dis} = 0.5$ (`model A') and $f_{\rm dis} = 0.1$ ('model B').
We will show, particularly in paper II, that 
our main results do not depend strongly on $f_{\rm dis}$, 
provided it is in this
range or smaller. We also note that in self-consistent simulations of
halo formation, few subhaloes will be resolved with sufficiently many
particles to trace their survival down to these levels, so we should not
be surprised if they resemble our model results for larger values of
$f_{\rm dis}$. The fraction of the original mass within 
$f_{\rm dis}\,r_{\rm bind,0}$ will of course depend on the
initial concentration of the system; this dependence is shown
in Fig.\ \ref{nfig:8} for the NFW profile, the Moore profile 
and the TN profile, for $f_{\rm dis} = 2.0$, $f_{\rm dis} = 0.5$, 
and $f_{\rm dis} = 0.1$.

\subsubsection{Implications}
In most previous semi-analytic treatments of subhalo evolution, dynamical
friction was assumed to be the main process responsible for the
destruction of substructure. 
For the orbits considered here, the time-scale for this process to occur is:
\begin{equation}
t_{\rm dis} = 1.2\,e\,{{(M_{\rm h}/M_{\rm s})}\over{\ln(M_{\rm h}/M_{\rm s})}}\,\epsilon^{0.4}\,{P_{\rm vir}\over{2\,\pi}}\,,
\label{eq:tcolpi}
\end{equation}
(Colpi, Mayer and Governato 1999), 
where $M_{\rm h}$ is the mass of the main halo, $M_{\rm s}$ 
is the initial mass of the subhalo, and $\epsilon$ is the initial circularity 
of the satellite's orbit. The factor $e > 1$ 
in this formula corrects for mass loss, which in this picture {\it increases}
the disruption time, by reducing the mass of the satellite and thus the
dynamical friction it experiences. As pointed out by Taffoni et al.\ (2002),
since $P_{\rm vir}$ is comparable to the Hubble time, this infall time
will be extremely long (hundreds or thousands of Hubble times) for all but the 
most massive satellites. On the other hand, we saw above that a typical 
satellite will be stripped of 25--45 per cent of its mass after each
orbital period, and that for many systems, this repeated mass loss 
may eventually result in their complete disruption. 

\begin{figure}
  \centerline{\psfig{figure=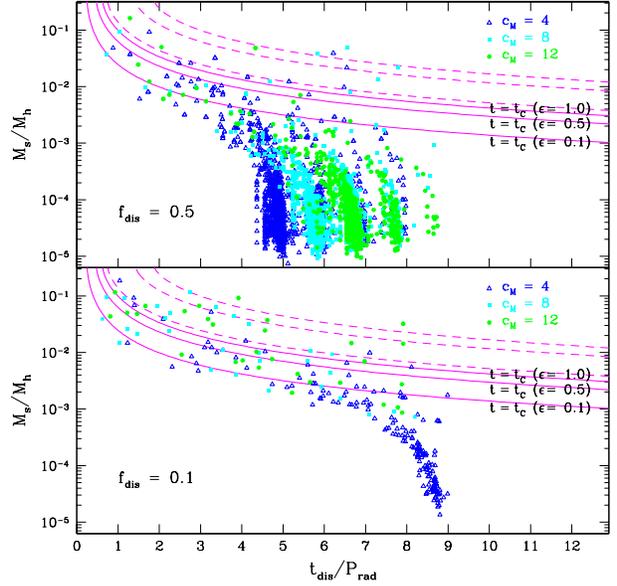,width=1.0\linewidth,clip=,angle=0}}
  \caption[]{Disruption times for satellites on a distribution of orbits,
in static halo with a Moore density profile of concentration $c_{\rm M} = 10$,
including dynamical friction. 
The point types indicate the concentration of the satellites. The smooth 
curves indicate the orbital decay time, as estimated from equation 
(\ref{eq:tcolpi}), for $e = 1$ (solid lines) and $e = 3$ (dashed lines),
and $\epsilon = 1.0$, 0.5 and 0.1 (from top to bottom). 
The top panel shows results
for $f_{\rm dis} = 0.5$; the bottom panel shows results for 
$f_{\rm dis} = 0.1$. Note that there are no satellites in these trees 
that have merged more than 9 radial periods previously, 
so we cannot measure disruption
rates beyond $t_{\rm dis}/P_{\rm Rad} = 9$.}
\label{nfig:9}
\end{figure}

Fig.\ \ref{nfig:9} shows the disruption times, that is the times after which
systems have been stripped down to 
$M_{\rm bind} = M(<f_{\rm dis}\,r_{\rm bind, 0})$, in a potential with a Moore
profile of concentration $c_{\rm M} = 10$, for satellites of concentration
4, 8 or 12 (different point types), on a representative distribution of 
orbits (see section \ref{subsec:2.3}) and including dynamical friction,
for $f_{\rm dis} = 0.5$ (model `A', top panel) and 
$f_{\rm dis} = 0.1$ (model `B', bottom panel). 
The smooth curves indicate the orbital decay
time calculated from equation (\ref{eq:tcolpi}), for $e = 1$ (solid lines) 
and $e = 3$ (dashed lines).  

We see that while the disruption time scales with mass and is comparable to 
the orbital decay time for the most massive systems 
($M_{\rm s}/M_{\rm h} \ga 0.01$), for less massive systems it becomes roughly 
independent of mass. In model `A' (top panel), most 
low-mass systems are disrupted due to repeated mass loss after 5--8 
pericentric passages, depending on their concentration, as 
expected from the mass-loss rates and critical mass fractions given above. 
Note that there are no satellites in these trees 
that have merged more than 9 radial periods previously, 
so we cannot measure disruption
rates beyond $t_{\rm dis}/P_{\rm Rad} = 9$.
From the results in the top panel, however, we expect
systems to disrupt after 9--12 orbits in model `B' (bottom panel). 
We conclude that while the dynamical friction time
derived by Colpi et al.\ (1999) is quite accurate for massive satellites,
with $e \simeq 2$, a proper description of mass loss and disruption
is essential to determining the evolution of less massive substructure.

\subsection{Summary}

In this section, we have considered the evolution of subhaloes in the static
potential generated by a larger system. We have chosen density profiles and 
orbital parameters which should be representative of the halo mergers that 
occur in hierarchical structure formation, and which therefore provide 
a well-defined set of initial conditions for studying the dynamics of halo 
substructure. We find that most mass loss occurs as a subhalo passes 
through the pericentre of its orbit, particularly for the radial orbits 
typical in cosmological settings. In general, we predict that
subhaloes should lose about 25--45 per cent of their 
mass for each pericentric passage, with the amount of mass loss tending to 
increase as the orbit becomes more radial. After a number of pericentric 
passages, this repeated mass loss may lead to complete disruption, although 
further simulations are required to confirm exactly when this takes place. 
In TB01 we also found that dynamical friction is also strongest close to the 
pericentre of the orbit.
Thus the overall dynamical state of a subhalo will depend principally on 
the number of pericentric passages, or the number of orbits (in the sense
of radial oscillations), that it has spent 
in its parent halo. In a static halo, this is simply 
$n_{\rm o} = \Delta\,t/P_{\rm rad}$, where $\Delta\,t$ is the time elapsed 
since the satellite first crossed the virial radius of the parent and 
$P_{\rm rad}$ is the (fixed) radial period defined above.

In a cosmological setting, the radial period of an orbit will vary if 
the mean density within the orbit changes. In particular, the radial 
period at the virial radius will increase roughly proportionally to time, 
as the density inside the virial radius decreases, and even orbits in the 
inner parts of a halo may develop longer periods if major mergers 
rearrange material in the halo sufficiently to reduce its mean central 
density. Subhaloes merging into the main system will have 
concentrations and densities that depend on their merger epoch. Furthermore, 
they may also contribute their own substructure, which has already 
experienced mass loss in earlier stages of the merging hierarchy, to
the main system. Thus some of the patterns seen 
in this section may be obscured in more general situations. 
In the next section we will 
consider subhalo evolution in a realistic system, 
that includes all of these effects. We will show in particular that 
satellite properties still correlate strongly with 
$\Delta\,t/P_{\rm rad}$, provided we use the radial period at the
time when the satellite {\it first crossed} the virial radius to estimate how
many times it has orbited within the larger system.

\section{Evolution of substructure in a realistic system}\label{sec:dynamic}

To study the evolution of substructure in a realistic halo, we have to take 
into account the changing mass and structural properties of the main system,
as well as any previous evolution of its subcomponents at earlier stages of 
the hierarchical merging process. The changing properties of the main system
can be derived using the semi-analytic methods described in sections 
\ref{subsec:2.1}--\ref{subsec:2.2}. 
Specifically, we can generate a set of representative mass accretion 
histories, that is functions $M(z)$, for haloes with a given mass at the 
present 
day, using extended Press-Schechter methods, and pick one of these to
represent the main system. The spherical collapse model
then gives us the virial radius $r_{\rm vir}$ and circular velocity 
$V_{\rm c}$ of the main system at each redshift, while the ENS01 relations 
give us its concentration $c$ and thus its scale radius $r_{\rm s}$.

The spectrum of infalling satellites is provided by the merger tree,
and their initial structural properties can be specified in the same way
as for the main halo, as described previously. 
Higher-order substructure, that is the substructure within
infalling objects, is more complicated to deal with. This substructure
may lose mass and be disrupted in a smaller parent halo higher up the
merger tree before it ever reaches the main system, complicating the picture
considerably. Our strategy, as explained in the introduction, will be to 
develop a simplified description of the evolution of systems in the main 
halo, and then apply this to the side-branches of the merger tree, adjusting 
the model parameters iteratively to achieve a consistent solution that
predicts the same evolution in the main branch as is assumed in
generating the underlying merger tree. 
In what follows, we will show results for merger trees where we 
have already performed this iteration, so the adopted parameters are
self-consistent; the process required to achieve this convergence will
be explained in section \ref{sec:pruning}.

\subsection{Orbital evolution}

Since period of radial oscillations at the virial radius of the main halo,
$P_{\rm rad}$, changes with time as it grows, we will consider the evolution 
of satellites within this system in terms of $P_{\rm rad}(z_{\rm m})$, 
the radial period the main halo at the time they first crossed its virial 
radius. Specifically, if a satellite merges at $z_{\rm m}$ and then spends
an interval of time $\Delta t$ in the main halo, we will use 
$\Delta t/P_{\rm rad}(z_{\rm m})$ as an estimate of the number of radial
oscillations it has undergone. This should be reasonably accurate
until the mean density interior to its orbit changes, as may happen in a 
major merger. For brevity, we will write this quantity as 
$\Delta t/P_{\rm rad}$ in what follows, $P_{\rm rad}$ implicitly being
measured at $z_{\rm m}$.

\begin{figure}
  \centerline{\psfig{figure=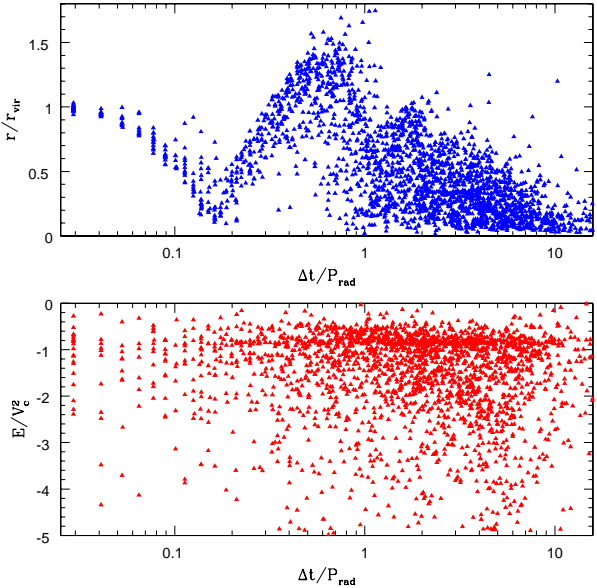,width=1.0\linewidth,clip=,angle=0}}
  \caption[]{(Top panel) Radial positions of the satellites from a large set 
of trees, evolved with model A ($f_{\rm dis} = 0.5$), versus
$\Delta\,t/P_{\rm rad}$. For clarity only one randomly chosen member of every 
dynamical group has been plotted. (Bottom panel) Orbital energy of the same 
objects, in units of the present-day circular velocity of the main halo 
squared, as a function of $\Delta\,t/P_{\rm rad}$.}
\label{nfig:10}
\end{figure}

Because we are assuming that the initial orbital energy of
infalling satellites depends only on the epoch at which they merge,
and the mass of the main halo at this time, we expect the radial coordinate 
of satellites to show strong correlations with $\Delta\,t/P_{\rm rad}$.
Fig.\ \ref{nfig:10} shows the radius and orbital energy of satellites from
many different merger trees, plotted as a function of 
$\Delta\,t/P_{\rm rad}$. (For clarity, we have only plotted
one randomly chosen member of every dynamical group -- see section 
\ref{subsec:5.3}). In the top panel, we see that satellites with recent merger
times are close to the virial radius, while successively older systems 
have reached the pericentre of their orbits, are moving back out to apocentre, 
or are on their second or third
orbit. In general, old systems are concentrated at smaller radii than
recently accreted ones (in fact the scatter in orbital apocentre for the
older systems is due mainly to the velocity dispersion given to dynamical 
groups, as discussed in section \ref{subsec:5.3}). 
The strong correlation between radius and merger 
epoch should produce patterns in substructure that are easily visible as 
coherent clumps or shells in individual haloes. We will discuss 
the grouping of substructure further in section \ref{subsec:5.3}
and in paper II.

There is 
also a slight correlation between $\Delta\,t/P_{\rm rad}$ and the orbital 
energies of satellites. As the bottom panel in Fig.\ \ref{nfig:10} shows, 
the mean orbital energy
decreases slightly with increasing $\Delta\,t/P_{\rm rad}$, indicating that the
oldest satellites within a halo are somewhat more bound than those 
that have merged more recently. We note, however, that this trend depends on
the disruption criterion assumed. We will discuss the effects of disruption
on the orbital energy distribution of substructure in detail in paper II.

In summary, since the orbits of recently accreted satellites are both more 
radially extended and less tightly bound, we conclude that these satellites 
will be the first to be stripped out of groups when they fall into 
larger haloes (ignoring for the moment the effects of dynamical friction
on the most massive satellites). 
In the pruning model discussed in section \ref{sec:pruning},
we will therefore use the quantity $\Delta\,t/P_{\rm rad}$ to 
distinguish between subhaloes that remain associated with their parent system,
and those that are stripped off, and should thus treated independently in the 
next level of the merger hierarchy.

\subsection{Mass loss}

\begin{figure}
  \centerline{\psfig{figure=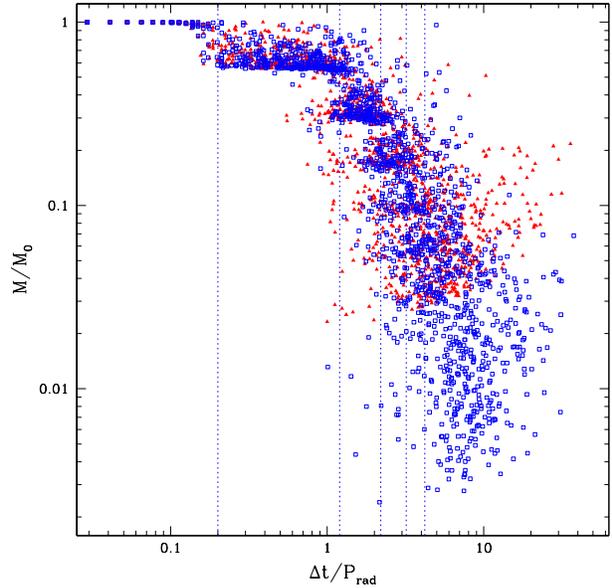,width=1.0\linewidth,clip=,angle=0}}
  \caption[]{The mass of individual satellites, as a fraction of their 
mass at the time of infall, versus $\Delta\,t/P_{\rm rad}$. Filled triangles
are for model A ($f_{\rm dis} = 0.5$), and open squares 
are for model B ($f_{\rm dis} = 0.1$). The vertical lines indicate the 
approximate time of successive pericentric passages.
}
\label{nfig:11}
\end{figure}

Given the results of section \ref{sec:static}, the average amount of mass 
loss for subhaloes should also depend strongly on $\Delta\,t/P_{\rm rad}$.
Fig.\ \ref{nfig:11} shows the bound mass fraction of individual satellites, 
that is their bound mass as a fraction of its value at the time of infall 
into the main system, versus $\Delta\,t/P_{\rm rad}$, for model A
($f_{\rm dis} = 0.5$ -- filled circles) and model B ($f_{\rm dis} = 0.1$ -- 
open squares). For clarity we have only plotted one satellite per
dynamical group, as in Fig.\ \ref{nfig:10}. 
The lines indicate the approximate time of successive pericentric passages.
We see that
the pattern evident in Fig.\ \ref{nfig:6} is reproduced, although with
more scatter and greater mass loss in some cases, due to the different 
orbits and concentrations considered and the effects of dynamical friction 
on the pericentric radius. Overall, the average bound mass fraction 
decreases for the first five or six radial periods, but increases slightly
thereafter, as the only systems surviving at this point have relatively
circular orbits or are more concentrated, raising the average. 

\subsection{Disruption}

\begin{figure}
  \centerline{\psfig{figure=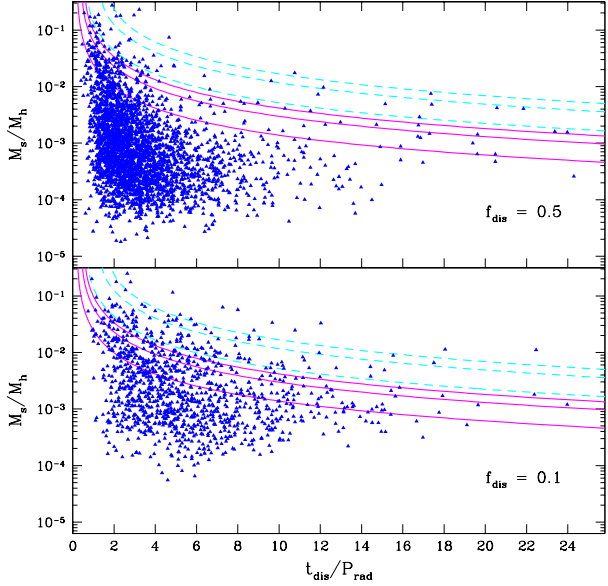,width=1.0\linewidth,clip=,angle=0}}
  \caption[]{As Fig.\ \ref{nfig:9}, but in realistic systems. The top panel 
shows the disruption time in units of the radial period at the time
when the satellite first merges, for model A ($f_{\rm dis} = 0.5$); 
the bottom plot shows the disruption times for model B ($f_{\rm dis} = 0.1$). 
The curves show the orbital decay time for $e = 1$ (solid) and $e = 3$ 
(dashed), and for $\epsilon = 1.0$, 0.5 and 0.1, as in Fig.\ \ref{nfig:9}.}
\label{nfig:12}
\end{figure}

Fig.\ \ref{nfig:12} shows disruption times for subhaloes 
in a realistic halo, where
$P_{\rm rad}$ changes with the virial density, that is roughly 
proportionally to $t$, and many of the infalling haloes are higher-order,
that is to say they have already merged with earlier systems before
their final merger with the main halo considered here. 
(We correct for mass loss due to this previous evolution self-consistently,
as explained in section \ref{subsec:5.2}.) The lines show
the estimated orbital decay time, as in Fig.\ \ref{nfig:9}.
We see that for a realistic halo, disruption times show more scatter
and the grouping at specific pericentric passages seen in Fig.\ \ref{nfig:9}
is obscured. On average, however, systems are disrupted earlier than in the 
static case, after
$\simeq$ 1--8 pericentric passages for model A ($f_{\rm dis} = 0.5$),
or $\simeq$ 2--15 pericentric passages for model B ($f_{\rm dis} = 0.1$). 
This is due to prior mass loss in earlier
stages of the hierarchical merging process, which reduces the effective
concentration of satellites, destabilising them. 

\begin{figure}
  \centerline{\psfig{figure=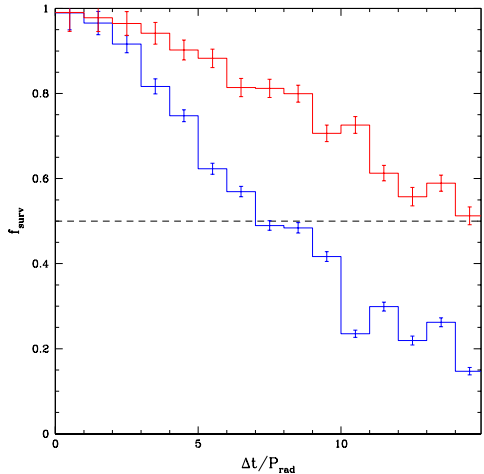,width=0.8\linewidth,clip=,angle=0}}
  \caption[]{The fraction of the subhaloes that have survived disruption, 
as a function of the number of radial periods they have spent in the main
system. The lower histogram is for model A and the upper histogram is for
model B.}
\label{nfig:13}
\end{figure}

Fig.\ \ref{nfig:13} shows the fraction of systems which have survived 
disruption, as a function of $\Delta\,t/P_{\rm rad}$. 
For model A ($f_{\rm dis} = 0.5$), the disruption probability reaches
50 per cent after 8 pericentric passages,
while for model B ($f_{\rm dis} = 0.1$), the disruption probability 
reaches 50 per cent after roughly 15 pericentric passages.

\begin{figure}
  \centerline{\psfig{figure=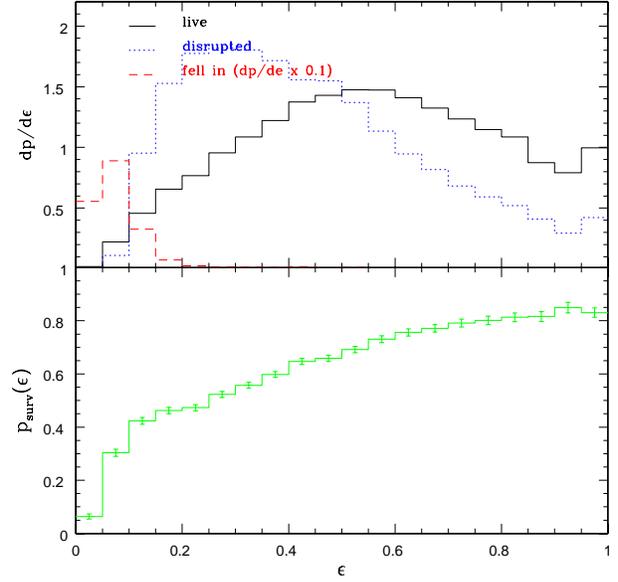,width=1.0\linewidth,clip=,angle=0}}
  \caption[]{(Top panel) the circularity distributions 
of surviving and disrupted satellites, as well as those that have fallen in,
for model A ($f_{\rm dis} = 0.5$).
(Bottom panel) the fraction of satellites surviving at the present day,
as a function of their 
initial circularity.}
\label{nfig:14}
\end{figure}

So far we have not discussed the dependence of the disruption time
on orbital circularity. Generally, disruption occurs sooner 
for more radial orbits, 
especially for the most extreme radial orbits. Fig.\ \ref{nfig:14} shows 
the distributions of initial circularity $\epsilon$ for surviving 
and disrupted satellites, in a set of trees evolved with 
model A ($f_{\rm dis} = 0.5$). In addition, we plot satellites that 
have `fallen in', that is objects whose orbits have taken them to 
pericentres we cannot resolve, below 1 per cent of $r_{\rm vir}$. 
(Naturally, these systems are on very radial orbits.) Since we cannot 
follow their subsequent evolution with any accuracy, we take them 
to be disrupted, as mentioned in section \ref{subsec:3.2}. 
Apart from this, there is also a clear shift between the surviving 
and disrupted distributions. In the bottom panel, we show the 
fraction of satellites surviving as a function of their initial 
circularity.  

\begin{figure}
  \centerline{\psfig{figure=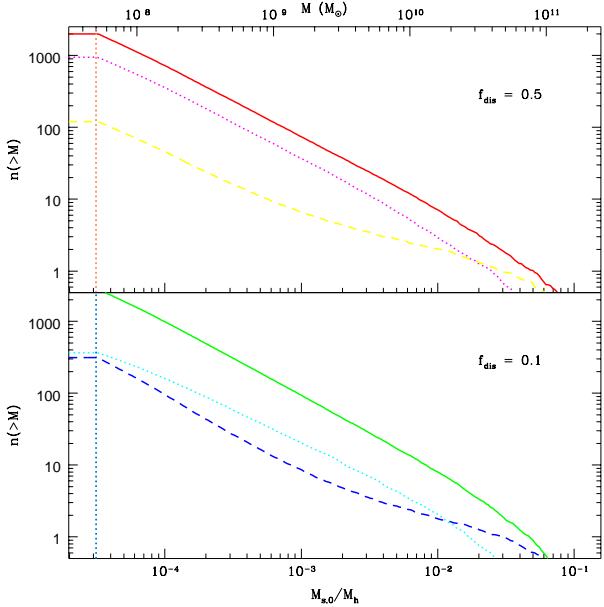,width=1.0\linewidth,clip=,angle=0}}
  \caption[]{Cumulative initial mass functions for all the subhaloes that merge with 
a main system between $z = 0$ and $z = 30$, averaged over many merger trees, 
for model A ($f_{\rm dis} = 0.5$, upper panel) and model B 
($f_{\rm dis} = 0.1$, lower panel). The solid line shows the mass function 
of surviving haloes, while the dotted and dashed lines show the mass functions 
of the subhaloes that have been disrupted by repeated mass loss, and those 
that have been disrupted by falling into the centre of the potential, 
respectively. In each case we plot the cumulative number of subhaloes
with more than some {\it initial} mass which wind up in that final end state.}
\label{nfig:15}
\end{figure}

Finally, we can consider the disruption rate as a function of the initial
mass of the satellite. Fig.\ \ref{nfig:15} shows cumulative initial
mass functions
for all the subhaloes that merge with a main system between $z = 0$
and $z = 30$, down to a mass resolution of $5\times 10^{7}\, {\rm M}_{\odot}$,
averaged over a large set of merger trees, for model A ($f_{\rm dis} = 0.5$, 
upper panel), and model B ($f_{\rm dis} = 0.1$, lower panel). 
The solid lines show the mass functions of surviving haloes, while the 
dotted and dashed lines show the mass functions of subhaloes disrupted by 
mass loss and subhaloes
disrupted by having fallen into the centre of the potential, respectively.
(In each case we plot the cumulative number of subhaloes with more than some
{\it initial} mass, since the 
the disrupted systems have no bound mass left at $z = 0$.)

Since the ratio of the solid and dotted curves is roughly constant,
we infer that the rate of disruption due to mass loss is approximately 
independent of mass. Approximately half of all systems are disrupted in
model A, whereas only a quarter are disrupted in model B (there is more
variation with mass in this case as well). The second disruption rate, 
due to objects falling into the centre of the potential,
is also roughly independent of mass at low masses; at high masses 
($M/M_{\rm vir,0} \ga 2\times 10^{-3}$) the effect of 
dynamical friction is evident from the change the slope of the 
dashed curves. Only 5--10 per cent of low-mass systems are disrupted
at the centre of the potential, whereas almost all of the massive ones
are. 

\subsection{Summary}

Having examined the dynamical evolution of substructure in a realistic
halo, we can conclude on some general features of subhalo dynamics.
First, the main dynamical properties of subhaloes are strongly
correlated with the quantity $\Delta\,t/P_{\rm rad}$ (where $P_{\rm rad}$
is evaluated at the time when the subhalo first crosses the virial radius of 
the main system), which provides 
an estimate of how many radial oscillations they have undergone in 
the larger system. Younger subhaloes (those with lower values of 
$\Delta\,t/P_{\rm rad}$) are on less bound and more extended orbits 
within their parent haloes. The average amount of mass loss, and by 
implication the degree of tidal stripping, is also strongly correlated 
with $\Delta\,t/P_{\rm rad}$, as is the fraction of systems that
have been disrupted. Disruption also occurs faster on radial orbits,
or for more massive satellites. In the next section we will put this 
information
together to suggest a way of treating higher-order substructure in merger
trees.

\section{A general method for pruning merger trees}\label{sec:pruning}

When a group of galaxies merges with a larger cluster, the
individual galaxies should remain on correlated orbits for some time. 
Over time, tidal forces and encounters within the cluster will 
strip the most loosely bound members from the group, so that
fewer and fewer objects remain closely associated. After many orbits 
in the cluster, only tightly bound galaxy pairs will remain associated. 
Similarly, when a small halo merges into a larger one,
its substructure may remain associated, if it is tightly bound,
or may be stripped off, if it is loosely bound. Since we expect
older substructure to be more tightly bound, and possibly even
disrupted, within the infalling halo, we can assume that the 
substructure stripped off it when it merges into a larger system
is preferentially `younger' (in the sense that it has merged into the halo
more recently). In what follows, we will outline a pruning method based on 
this idea.

\subsection{Self-similar pruning}

\subsubsection{How much to prune?}

Let us consider the fate of higher-order substructure in a merger tree, 
say subhaloes within a parent halo that merges with an even larger 
system. As the parent falls into the large system, it will lose some fraction
of its mass, say $\Delta M$, and will be stripped from the outside in. This 
stripped mass should include subhaloes, since they too can be accelerated 
away from the parent's orbit by the tidal field of the larger system. 
The kinematic distribution of subhaloes within the parent may be 
biased with respect to its smoothly distributed mass (for instance
if subhaloes closer to the centre of the parent have been preferentially 
disrupted), but we can estimate what fraction $f_{\rm st}$ of the subhaloes 
are stripped off by determining how they are distributed in radius
within the parent, and thus what fraction of the subhaloes are situated
in the outer region containing $\Delta M$ of the total mass. 
If we perform this calculation for $\Delta M = \overline{\Delta M}$, the 
average mass fraction lost by all subhaloes in the main system up to the 
present day, then we will have an estimate of 
the average fraction of substructure stripped off all systems.

\begin{figure}
  \centerline{\psfig{figure=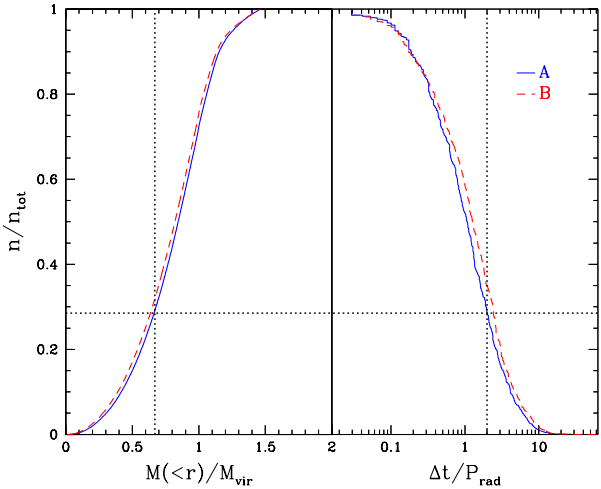,width=1.0\linewidth,clip=,angle=0}}
  \caption[]{(Left-hand panel) Fraction of systems interior to a given
radius, plotted versus the fraction of the total mass of the
main system contained within that radius. The two curves are for
model A ($f_{\rm dis} = 0.5$, solid line) and 
model B ($f_{\rm dis} = 0.1$, dashed line). (Right-hand panel) 
Fraction of systems which have spent more than a given number of orbits 
in the system, as a function of $n = \Delta\,t/P_{\rm rad}$. 
The lines indicate how various fractions of the subhalo population
correspond to one another in model A (see text).}
\label{nfig:16}
\end{figure}

The left-hand panel of Fig.\ \ref{nfig:16} shows the average fraction of 
subhaloes within a given radius, as a function of the fraction of the mass 
inside that radius, for a large set of model haloes. The two curves 
are for model A ($f_{\rm dis} = 0.5$, solid line) and model B 
($f_{\rm dis} = 0.1$, dashed line). The satellites are concentrated 
closer to the centre of the main system in the latter case, but the offset 
between the two distributions is small. For model A, the average mass 
fraction lost by subhaloes in the main system is $\overline{\Delta M} = 0.33$.
The dashed lines indicate that 72 per cent of the subhaloes in a typical halo
will lie outside the corresponding radius, that is 72 per cent of 
all subhaloes lie in the region containing the outermost 33 per cent of 
the mass. Thus in model A, $f_{\rm st} = 0.72$ on average.

\subsubsection{Which systems to prune?}

In section \ref{sec:dynamic}, we saw that younger systems are typically
on larger, 
less bound orbits. If we assume that the subhaloes most recently accreted
by the parent will be the first to be stripped off,
then we can choose a value $n_{\rm o}$ such that the fraction of 
subhaloes 
that have spent less than $n_{\rm o}$ orbits in the parent is equal 
to $f_{\rm st}$.
The right-hand panel of Fig.\ \ref{nfig:16} shows the relative number
of subhaloes in a merger tree with $\Delta\,t/P_{\rm rad} > n_{\rm o}$, 
as a function of $n_{\rm o}$. We see that
for $f_{\rm st} = 0.72$, $n_{\rm o} = 2.0$, that is 72 per cent of the 
subhaloes in an average halo have spent less than 2 radial periods in 
their parent system. These are roughly the systems that will be stripped 
off if the parent loses 33 per cent of its mass.
By assuming the most recently acquired haloes are stripped off, we
can specify our
pruning algorithm in terms of a single parameter,
say $n_{\rm o}$. For a given value of $n_{\rm o}$, the corresponding
values of $\overline{\Delta M}$ and $f_{\rm st}$ can be determined from the
distributions in Fig.\ \ref{nfig:16}, or alternately $n_{\rm o}$
and $f_{\rm st}$ can be determined from $\overline{\Delta M}$. 

We also saw that orbital properties and disruption rates of satellites 
depend on the initial circularity of their orbit. We could include 
this dependence explicitly in our stripping algorithm, but since successive
mergers at each level of the hierarchy destroy all information about a 
satellite's orbit in earlier systems, we do not lose any precision by 
averaging over results for different circularities. We should correct
for the orbital decay produced by dynamical friction, however;
as shown in Fig.\ \ref{nfig:12}, the time-scale for this process will
be comparable to, or shorter than, 2 radial periods for satellites with
masses of $\sim 10^{-2} M_{\rm vir}$ or more, and a substantial fraction 
of the most massive satellites will be disrupted completely due to
orbital decay (cf.\ Fig.\ \ref{nfig:15}). Thus, we revise our pruning
criterion slightly, 
stripping from their parent only those systems that have spent
less than $n_{\rm o}$ radial periods in the parent halo, {\it and} have
orbital decay times longer than $n_{\rm o}\,P_{\rm rad}$. 
(Based on our mass-loss
estimates, and Figs. \ref{nfig:9} and \ref{nfig:12}, we calculate the orbital
decay time using equation (\ref{eq:tcolpi}) with a value of $e = 2$.)

\subsubsection{Fixing parameters}

Once we have generated merger trees using initial estimates of 
$\overline{\Delta M}$, $f_{\rm st}$ and $n_{\rm o}$, we can evolve them and
measure new values for these parameters in the main trunk, where
we can follow subhalo evolution in detail. If we assume that the
evolution of substructure is self-similar in the main trunk and the 
branches, then iterating through this process will fix the values
of the parameters by self-consistency. For a reasonable choice of 
initial guesses the iteration converges quickly, and we find that for 
model A, $\overline{\Delta M} = 0.32 \pm 0.002$ (where the uncertainty is the 
uncertainty in the average, not the variance of the distribution), 
$f_{\rm st} = 0.71 \pm 0.003$ and $n_{\rm o} = 2.0 \pm 0.05$, 
while for model B, $\overline{\Delta M} = 0.335 \pm 0.002$, 
$f_{\rm st} = 0.677 \pm 0.003$ and $n_{\rm o} = 2.25 \pm 0.05$
These averages are determined for all systems over our resolution limit, 
irrespective of mass. In theory the actual values will depend on our
treatment of orbital decay due to dynamical friction, but in practice
the averages are completely dominated by low-mass systems, for which
dynamical friction is negligible. 

Thus we have established 
a non-parametric method for pruning the side-branches of 
merger trees. The method is approximative in several ways -- for instance 
we assume subhaloes are stripped off in a strictly first-in-first-out order. 
We also assume that the properties of substructure, specifically,
the average fraction of mass lost by subhaloes, the distribution of
their dynamical ages, and their spatial distribution, are self-similar in the 
main trunk and in the branches of the merger tree. Simulations show that
this is roughly true over a wide range of scales, from galaxy haloes to
the haloes of massive clusters, so it seems a reasonable approximation.
Overall, the method proposed here provides a simple and effective way of 
handling higher-order substructure in merger trees. 
In the next section, we will discuss how to implement the method in practice.

\subsection{Implementing the method}\label{subsec:5.2}

\begin{figure}
  \centerline{\psfig{figure=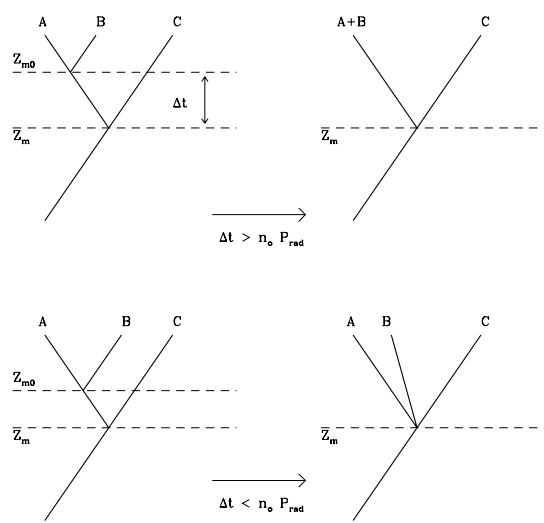,width=1.0\linewidth,clip=,angle=0}}
  \caption[]{A schematic illustration of the pruning process. If a satellite
`B' spends less than $n_{\rm o}$ radial periods in the halo of its parent `A'
(and if its orbit does not decay), 
then it is considered to be distinct object associated with `A' when `A' 
merges with a larger system `C' (bottom figures). 
Otherwise A and B are treated as a single system (top figures).}
\label{nfig:17}
\end{figure}

We can implement the pruning method outlined above as follows.
In each side-branch where there is higher-order
substructure, we determine for each subhalo the number
of orbits it has spent in its initial parent system, measured
by $\Delta t/P_{\rm rad}$ (where $P_{\rm rad}$ is evaluated 
at the time of the subhalo's original merger with its parent, 
as in section \ref{sec:dynamic}). We also calculate the orbital
decay time, from equation (\ref{eq:tcolpi}).
If $\Delta t/P_{\rm rad}$ is larger than some number of orbits 
$n_{\rm o}$, or if the orbital decay time is shorter than $\Delta t$, 
then we consider the system disrupted within its parent, or so tightly 
bound that it will not be stripped off its parent subsequently. 
If $\Delta t/P_{\rm rad}$ is less than $n_{\rm o}$ and the orbital 
decay time exceeds
$\Delta t$, then we assume the system will be stripped from its parent when
it merges with a larger halo. We treat the system as a distinct subhalo which
merges into the main tree at the same time its parent does, 
on an associated orbit. Fig.\ \ref{nfig:17} illustrates this pruning
process schematically.
To fix $n_{\rm o}$, we determine the average amount of mass lost by subhaloes
in the main system $\overline{\Delta M}$, and derive the corresponding fraction
of subhaloes stripped off along with this mass, $f_{\rm st}$, and number of
orbits $n_{\rm o}$ corresponding to $f_{\rm st}$, as explained in the 
previous section.

There are one or two other details to sort out in this model. First, we
must determine what structural parameters to use for higher-order
substructure. The concentration of a higher-order subhalo 
should reflect its original mass and relative age 
when it first fell into it a parent halo, since
it would have stopped growing at this point. These quantities
can be determined from the merger tree. While in
its parent system (before the merger with the main system), it
would also have evolved as described in section \ref{sec:dynamic}, losing
a fraction $\Delta M_{\rm st}$ of its mass. We can take $\Delta M_{\rm st}$
to be the average fraction of mass lost by subhaloes that have spent
less than $n_{\rm o}$ orbits in the main halo (note that this will be less
than the average mass-loss fraction for all subhaloes, $\overline{\Delta M}$).
Individual subhaloes may be passed on through many levels of the hierarchy,
losing mass repeatedly this way until they are disrupted.

\begin{figure}
  \centerline{\psfig{figure=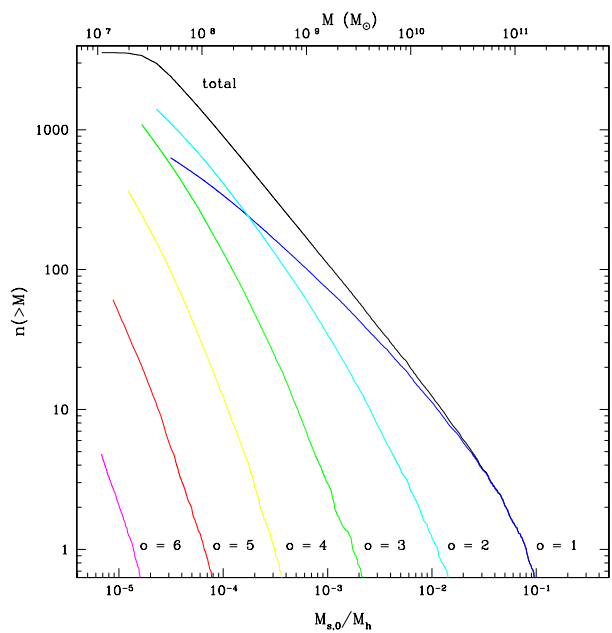,width=1.0\linewidth,clip=,angle=0}}
  \caption[]{The average cumulative mass function of all subhaloes merging 
into a system with a present-day mass of 
$M_{\rm h} = 1.6\times 10^{12} {\rm M}_{\odot}$ 
between $z = 0$ and $z = 30$ (top line), as well as
the contributions from haloes of successively higher order (lower lines,
with order increasing from top to bottom).}
\label{nfig:18}
\end{figure}

With these adjustments made, we now have a full model of halo
evolution which, beyond the free parameters $\Lambda_{\rm s}$ 
and $\epsilon_{\rm h}$ 
used to describe the evolution of single satellites and fixed by
comparison with high-resolution simulations in TB01,
has no other major free parameters. The only remaining parametric freedom
is in the choice of the disruption criterion $f_{\rm dis}$, and we will
show in paper II that this only has a minor effect
on the subhalo mass function for reasonable choices of $f_{\rm dis}$.
The pruning process can be iterated for successively higher-order branches, 
producing high-order haloes that may have survived many merging episodes. 
Fig.\ \ref{nfig:18} shows the average cumulative mass 
functions of subhaloes of 
various different orders that merge into a single halo of mass 
$M_{\rm vir,0} = 1.6\times 10^{12} {\rm M}_{\odot}$, 
(that is the cumulative distribution
of their initial masses, when they first merge with the main system).
The effect of multiple stripping
is clear in the decreasing masses for successively higher-order haloes. 
We also see that higher-order 
substructure helps to generate a scale-invariant cumulative mass 
function at low masses (top-most curve). Even at a mass of 
$M_{\rm s} = 10^{-3} M_{\rm h}$, higher-order
substructure accounts for half the subhaloes in a typical system. 
In paper II we will show that this contribution is required to match 
the halo mass functions measured in numerical simulations. 

\begin{figure}
  \centerline{\psfig{figure=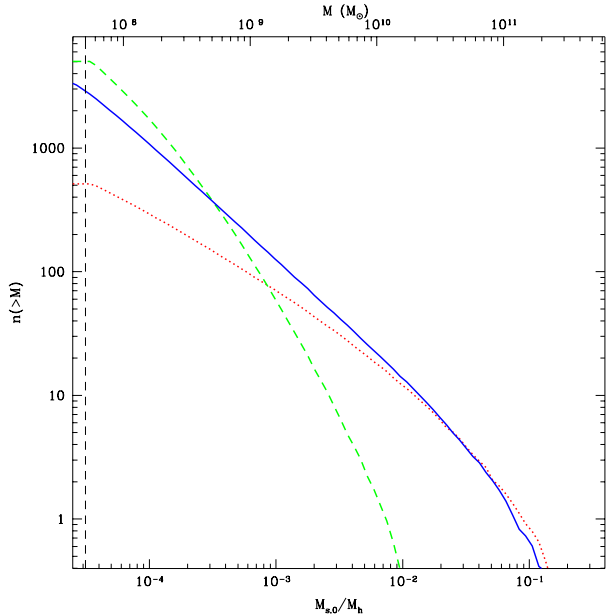,width=1.0\linewidth,clip=,angle=0}}
  \caption[]{The initial mass function of merging haloes, as in
Fig.\ \ref{nfig:18}, assuming no pruning, pruning with $n_{\rm o} = 2$, and 
pruning with very large $n_{\rm o}$ (dotted, solid and dashed curves 
respectively).}
\label{nfig:19}
\end{figure}

Pruning with $n_{\rm o} \simeq 2$ also produces the right slope for 
this power-law tail to the subhalo mass function. 
Fig.\ \ref{nfig:19} shows input
mass functions for basic merger trees where the side-branches are treated 
as single, monolithic objects (dotted curve), trees pruned 
with $n_{\rm o} = 2.0$ (solid curve), and trees 
where $n_{\rm o}$ is very large, so that all substructure is counted 
separately down to 
the resolution limit of the trees (dashed curve). Pruning transfers some of the
material in low-order subhaloes to smaller, higher-order subhaloes, reducing
the masses of the former by a small amount while greatly increasing the number 
of the latter. The net effect on the input mass function is to lower its amplitude
slightly at the high-mass end, while steepening the slope considerably 
at the low-mass end (compare the solid and dotted curves 
in Fig.\ \ref{nfig:19}). In the limit of large $n_{\rm o}$, most of the mass
in large systems is decomposed into small systems close to the mass resolution
of the merger tree (dashed curve). Subsequent mass loss and 
disruption will modify these input mass functions, particularly 
at the high-mass
end, as discussed in paper II. Even without considering this subsequent 
evolution, however, the low-mass slope of the intermediate curve is 
fairly close to the value of 1.8--2.0 measured in simulations 
(Ghigna et al.\ 1998, 2000; Moore et al.\ 1999; Springel et al.\ 2001). 
Finally, higher-order substructure should also introduce 
orbital correlations in halo substructure.
We will consider this point in the next section.

\subsection{Dynamical groups}\label{subsec:5.3}

When a subhalo that has survived the pruning process merges into 
the main system alongside its parent, it should have displacements 
in position and velocity that reflect the size $r_{\rm vir,g}$, mass 
$M_{\rm g}$, and circular velocity $V_{\rm c,g}$ of the group it
falls in with. It is not completely clear how large to make these offsets, 
since on the one hand we are stripping systems preferentially from 
the outer parts of the group and the less bound orbits, but on the
other hand we cannot account for the self-gravity of the group in
our model. In any case, the details of the initial distribution will
no longer be relevant after the group has passed through pericentre
once and the orbits have been scattered. For simplicity, we will
choose an offset in radius picked from a uniform distribution truncated 
at $\Delta r = r_{\rm vir,g}$ (equivalent to an isothermal density
distribution within this radius), and an offset in velocity picked 
from a Maxwellian distribution of width $\Delta V = V_{\rm c,g}$,
with a cutoff at $2\,V_{\rm c,g}$.

These offsets correspond to the approximate spatial and velocity distributions
of substructure in hierarchically assembled haloes, so in effect we are simply
putting realistic haloes with multiple components into the main potential.
A slightly different interpretation of the offsets 
is that groups lose material, just as
as individual haloes do, into streams with a characteristic initial 
scale comparable to $r_{\rm vir,g}$ and a velocity dispersion comparable 
to $V_{\rm c,g}$ (see for instance Johnston 1998).
Of course, the short-coming of this method is that the subsequent
evolution of the group neglects its self-gravity, as mentioned previously,
and thus the total velocity dispersion of our haloes may be slightly larger 
than it should be. We will discuss this further in paper II.

\begin{figure}
  \centerline{\psfig{figure=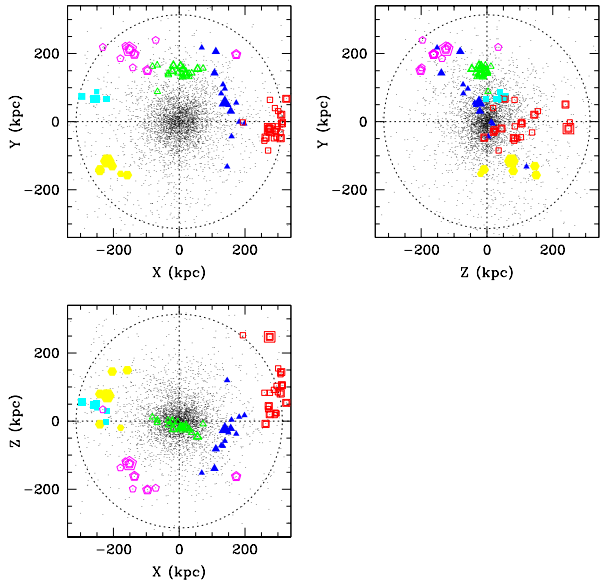,width=1.0\linewidth,clip=,angle=0}}
  \caption[]{The projected positions of the members of several young
groups, in a halo that has experienced many recent mergers. The large
symbols are group members, with point type indicating group membership
and point size giving an indication of 
mass; the small dots are the other subhaloes in the system.}
\label{nfig:20}
\end{figure}

Having included this description of orbital correlations in our model, 
we now have a 
way of studying groups of substructure within larger systems. 
Fig.\ \ref{nfig:20}, for instance, shows a few young groups in a halo that
has experienced many recent mergers. Similar substructure can be observed
both in galaxy clusters, particularly with the recent availability 
of high-resolution X-ray imaging and multi-object spectroscopy 
(e.g.\ Sun, Murray, Markevitch, \& Vikhlinin 2002; 
Mazzotta, Fusco-Femiano, \& Vikhlinin 2002; 
Berrington, Lugger, \& Cohn 2002; Belsole et al.\ 2002; Bardelli et al.\ 2002)
and even in the halo of Milky Way, where the Magellanic Clouds and some of 
the other satellites 
appear to be members of dynamical groups on associated orbits 
(Lynden-Bell 1976; see Binney 2001, and Palma, Majewski, \& Johnston 2002 
for recent references). There is the intriguing possibility
of reconstructing the merger history of our galaxy
using these observed groups, particularly if more accurate kinematic
information becomes available through interferometric satellites such as
{\it GAIA} (Binney 2001). We will consider the properties and evolution of
dynamical groups further in paper II.

\section{Summary}\label{sec:summary}

There is increasing evidence, both observational and theoretical, that
the dark matter haloes surrounding galaxies, groups and clusters contain
substructure on a wide range of scales. It is beyond current computational 
capabilities to determine the properties of this substructure directly.
To take full advantage of results from lensing observations, dark matter 
direct-detection experiments, and searches for a WIMP annihilation signal in 
the gamma-ray background, and to provide a robust platform for detailed
models of small-scale structure formation in general cosmologies, 
an analytic or semi-analytic extension to current numerical 
methods is required. A key problem in devising such a model is to determine
how much substructure to pass on from one level of the merger hierarchy 
to the next.

In this paper, we have applied the analytic model of satellite evolution
developed in TB01 to the study of merging haloes, using the
initial conditions for these mergers expected in a cosmological setting.
We find that several basic patterns characterise subhalo dynamics.
The main time-scale for subhalo evolution is the period for radial 
oscillations at the virial radius, 
$P_{\rm rad} \simeq 5\,r_{\rm vir}/V_{\rm c}$ (evaluated at the time
when the subhalo first crosses the virial radius of the main system),
since this is the period for successive 
pericentric passages in the satellite's orbit.
Around each pericentric passage, the subhalo loses mass through 
tidal stripping; for representative densities and density profiles
roughly 25--45 per cent of the remaining mass is lost on each successive
orbit, the exact fraction depending on the circularity of the orbit and 
the concentration of the satellite. After 5--10 orbits, systems may be 
disrupted completely by this repeated mass loss. Orbital decay 
due to dynamical friction is also an important factor in subhalo evolution, 
but only for the most massive subhaloes, those with a few per cent of the mass 
of the main system or more. 

Considering the average properties of subhaloes in realistic systems,
generated using semi-analytic merger trees, we find correlations
between the merger epoch of the subhalo, its average orbital radius,
and to a lesser extent its orbital energy. Based on this correlation, 
and on the observation that haloes are stripped of their mass from
the inside out (e.g.\ Hayashi et al.\ 2003), we suggest a method for
`pruning' merger trees, that is determining how much distinct
substructure should be passed on from one level of the merging hierarchy 
to the next. 

When haloes with substructure merge into larger
systems, they should lose mass just as simple haloes do. This mass 
will include some self-bound substructure, so tidal stripping should produce 
groups of independent subhaloes on orbits similar to that of the original
parent. To determine how much substructure to strip from a parent, we assume 
self-similarity in the merger tree. On average, a parent in a side-branch
should lose a fraction of its mass that can be determined directly from
the evolution of satellites in the main halo in our model. Assuming that this
mass is stripped from the outside in, and that the outer material
consists of the most recently accreted substructure, we determine
that subhaloes should be passed down the merger tree if they have
spent less than $n_{\rm o} \simeq 2$ orbits in their parent halo by 
the time it merges with a larger system, and that they should be subsumed 
into their parent otherwise. The critical number of orbits $n_{\rm o}$ is 
initially a free parameter in our model, but we can fix it iteratively by 
requiring that our model be self-consistent, that is by assuming the same 
average properties for subhaloes in the side-branches as are measured 
directly for subhaloes in the main trunk of the merger tree. 

Obviously, the ultimate value of our method rests on how well it reproduces 
the results of high-resolution simulations of halo substructure, for
standard cosmologies. In paper II, we will test our method by comparing 
to a series of simulations
of galaxy and cluster haloes. We will show that in general there is 
reasonable agreement between the two methods, but that we see a few 
discrepancies between the numerical and the semi-analytic results. 
We will argue that some of these discrepancies may actually be due 
to numerical effects in the simulations, and that overall, our 
method provides a reliable estimate of small-scale substructure
within dark matter haloes.

\vspace{-5mm}

\section*{Acknowledgements}

The authors wish to thank E. Hayashi, S. Ghigna, F. Governato, B. Moore, 
J. Navarro, T. Quinn and J. Stadel for providing data from their simulations 
for comparison with our model. We also wish to thank E. Hayashi, T.\ Kolatt, 
K. Moodley, J.\ Navarro, J. Silk, and S.\ White for helpful discussions. 
JET gratefully acknowledges the support of a postgraduate scholarship from the 
Natural Sciences \& Engineering Research Council of Canada (NSERC) during the 
initial stages of this work, and support from the Leverhulme Trust in the
latter stages. AB acknowledges support from NSERC through the Discovery
Grant Program. AB would also like to acknowledge the kind hospitality shown 
to him at CITA during the tenure of his CITA Senior Fellowship.

\end{document}